\documentclass[fleqn]{aa}
\usepackage{graphicx}
\usepackage{amssymb}
\usepackage{amsmath}
\usepackage{natbib}
\usepackage{mathrsfs}
\usepackage{ulem}
\usepackage{txfonts}
\usepackage{lmodern}
\usepackage{booktabs}
\usepackage{multicol}
\usepackage{cases}
\usepackage{color}
\usepackage{multirow,bigdelim}

\usepackage[switch]{lineno} 

\newcommand{\kepler}{\textit{Kepler}}

\def\m2s2{\,m$^{2}$\,s$^{-2}$} 

\newcommand{\vaisala}{Brunt-V\"ais\"al\"a}

\newcommand{\mesa}{\textsc{MESA}}

\newenvironment{itemize*}%
  {\begin{itemize}%
    \setlength{\itemsep}{1pt}%
    \setlength{\parskip}{1pt}}%
  {\end{itemize}}

\newcommand{\omg}{\langle\Omega\rangle_{\rm g}}
\newcommand{\omp}{\langle\Omega\rangle_{\rm p}}

\newcommand\T{\rule{0pt}{2.6ex}}
\newcommand\B{\rule[-1.2ex]{0pt}{0pt}}

\begin{document}
\title{Seismic evidence for near solid-body rotation in two \textit{Kepler} subgiants and implications for angular momentum transport}
\titlerunning{Seismic evidence for near solid-body rotation in two \textit{Kepler} subgiants}
\author{
S. Deheuvels\inst{1}
\and J. Ballot\inst{1} 
\and P. Eggenberger\inst{2}
\and F. Spada\inst{3}
\and A. Noll\inst{1}
\and J. W. den Hartogh\inst{4,5}
}

\institute{IRAP, Universit\'e de Toulouse, CNRS, CNES, UPS, (Toulouse), France
\and Observatoire de Gen\`eve, Universit\'e de Gen\`eve, 51 Ch. des Maillettes, CH-1290 Sauverny, Suisse
\and Max-Planck Institut f\"ur Sonnensystemforschung, Justus-von-Liebig Weg 3, 37077 G\"ottingen, Germany
\and Konkoly Observatory, Research Centre for Astronomy and Earth Sciences, Konkoly Thege 15-17, H-1121 Budapest, Hungary
\and ELTE Eötvös Loránd University, Institute of Physics, Budapest, Hungary
}

\offprints{S. Deheuvels\\ \email{sebastien.deheuvels@irap.omp.eu}
}


\abstract{Asteroseismic measurements of the internal rotation of subgiants and red giants all show the need for invoking a more efficient transport of angular momentum than theoretically predicted. Constraints on the core rotation rate are available starting from the base of the red giant branch (RGB) and we are still lacking information on the internal rotation of less evolved subgiants.}
{We identify two young \kepler\ subgiants, KIC8524425 and KIC5955122, whose mixed modes are clearly split by rotation. We aim to probe their internal rotation profile and assess the efficiency of the angular momentum transport during this phase of the evolution.}
{Using the full \kepler\ data set, we extracted the mode frequencies and rotational splittings for the two stars using a Bayesian approach. We then performed a detailed seismic modeling of both targets and used the rotational kernels to invert their internal rotation profiles using the MOLA inversion method. We thus obtained estimates of the average rotation rates in the g-mode cavity ($\omg$) and in the p-mode cavity ($\omp$).}
{We found that both stars are rotating nearly as solid bodies, with core-envelope contrasts of $\omg/\omp=0.68\pm0.47$ for KIC8524425 and $\omg/\omp=0.72\pm0.37$ for KIC5955122. This result shows that the internal transport of angular momentum has to occur faster than the timescale at which differential rotation is forced in these stars (between 300 Myr and 600 Myr). By modeling the additional transport of angular momentum as a diffusive process with a constant viscosity $\nu_{\rm add}$, we found that values of $\nu_{\rm add}>5\times10^4$~cm$^2$.s$^{-1}$ are required to account for the internal rotation of KIC8524425, and $\nu_{\rm add}>1.5\times10^5$~cm$^2$.s$^{-1}$ for KIC5955122. These values are lower than or comparable to the efficiency of the core-envelope coupling during the main sequence, as given by the surface rotation of stars in open clusters. On the other hand, they are higher than the viscosity needed to reproduce the rotation of subgiants near the base of the RGB.}
{Our results yield further evidence that the efficiency of the internal redistribution of angular momentum decreases during the subgiant phase. We thus bring new constraints that will need to be accounted for by mechanisms that are proposed as candidates for angular momentum transport in subgiants and red giants.}


\maketitle

\section{Introduction \label{sect_intro}}

Understanding how angular momentum (AM) is transported in stellar interiors is one of the main challenges faced by modern stellar astrophysics. Despite the important impact that rotation is known to have on stellar structure and evolution (see, e.g., \citealt{maeder09}), the mechanisms responsible for the evolution of rotation inside stars remain poorly understood. Current stellar evolution models generally include a prescription for hydrodynamically-induced AM transport through meridional circulation and shear instabilities (\citealt{zahn92}, \citealt{mathis04}, \citealt{mathis18}), but a growing body of evidence shows that they are far too inefficient at redistributing AM. They indeed fail to produce the nearly uniform rotation of the Sun, as revealed by helioseismology (\citealt{schou98}, \citealt{gough15}). They predict low coupling intensities between the core and the envelope of young stars, whereas short coupling timescales are needed to account for the surface rotation of stars in young clusters (\citealt{denissenkov10a}, \citealt{gallet13}). These models also drastically overestimate the core rotation rates of subgiants and red giants, as is detailed below.


Asteroseismology has given new momentum to this field by providing measurements of the internal rotation of stars of various masses, at different stages of their evolution. Most of them were obtained using the high-precision photometric data from the \kepler\ satellite (\citealt{borucki10}). In particular, asteroseismology has given us a rather clear view of how the internal rotation of low-mass stars ($M\lesssim2M_\odot$) varies along their evolution from the main sequence all the way to the core-helium-burning phase. Current observations seem to indicate that low-mass stars have only modest differential rotation during the main sequence. Beside the solar case, nearly uniform rotation was also found for solar-like main-sequence stars using \kepler\ data (\citealt{benomar15}, \citealt{nielsen15}). Probing the rotation in the deep cores of main-sequence stars would require to detect g modes, which is notoriously challenging for stars with convective envelopes (see, e.g., \citealt{appourchaux10}). \cite{fossat17} recently claimed to have found evidence for a fast rotating core in the Sun using a modulation of p modes that they interpret as caused by g modes. The validity of this result has however been questioned by several studies (\citealt{schunker18}, \citealt{appourchaux19}). After the end of the main sequence, the nonradial oscillation modes of subgiants and red giants develop a mixed character (they behave as p modes in the envelope and as g modes in the core). The detection of these modes with \kepler\ data has made it possible to precisely measure the core rotation. It was thus shown that the core spins up when stars reach the base of the red-giant-branch (RGB)  and that a significant radial differential rotation develops (\citealt{deheuvels14}, hereafter D14). This behavior is expected, considering that the core contracts while the envelope expends during this phase, but the seismically measured core rotation rates are in fact much lower than would be predicted if the layers were conserving their specific AM. As stars ascend the RGB, the core rotation rate remains roughly constant (\citealt{beck12}, \citealt{deheuvels12}, \citealt{mosser12b}, \citealt{gehan18}), despite the ongoing core contraction. All these observations point to the existence of an efficient redistribution of AM in subgiants and red giants.

The origin of the transport of AM in red giants remains a matter under much debate. It has been shown that purely hydrodynamical mechanisms of AM transport are much too inefficient to account for the core rotation of red giants (\citealt{eggenberger12}, \citealt{ceillier13}, \citealt{marques13}). \cite{mathis18} have proposed a revised modeling of the horizontal transport induced by shear instability, which is still too inefficient to reproduce asteroseismic observations. Other candidates have been considered for this additional transport of AM, such as internal gravity waves (IGW) excited at the bottom of the convective envelope either by turbulence (\citealt{fuller14}) or by plumes \citep{pincon16}. The latter could account for the rotation of young red giants near the base of the RGB (\citealt{pincon17}), but for more evolved red giants IGW are expected to damp before reaching the core and cannot produce the required coupling between the core and the envelope. Mixed modes themselves might efficiently transport AM for these stars (\citealt{belkacem15b}). Another candidate for the transport of AM in red giants is magnetism. The core of red giants could indeed be the seat of magnetic fields, either of fossil origin or the remnants of fields induced by a dynamo process in the convective core during the main sequence. The interaction of these fields with differential rotation could produce an efficient transport of AM (\citealt{maeder14}, \citealt{kissin15}). A magnetic transport of AM could also be induced by turbulence associated with magnetohydrodynamic (MHD) instabilities (\citealt{rudiger15}, \citealt{jouve15}, \citealt{fuller19}). So far, their impact on the rotation evolution of subgiants and red giants has been addressed only under the debated assumption of the Tayler-Spruit dynamo in radiative interiors (\citealt{spruit02}), and its revision based on a different saturation process by \cite{fuller19}. Comparisons with asteroseismic measurements have shown that this process is currently unable to correctly account for the internal rotation of subgiants and red giants (\citealt{cantiello14}, \citealt{denhartogh19}, \citealt{eggenberger20}, \citealt{denhartogh20}).

So far, seismic measurements of the core rotation are available starting from the base of the RGB (D14), near the transition between subgiants and red giants (\citealt{mosser14}). We are still lacking observational constraints on the internal rotation of less evolved subgiants. Such measurements would be particularly helpful to understand the transition between the nearly-uniform rotation of main-sequence stars and the growing core-envelope contrast at the base of the RGB. It is also important to stress that young subgiants undergo a strong forcing of differential rotation owing to the structural adjustment at the main sequence turnoff, especially for stars that had a convective core during the main sequence (\citealt{eggenberger19}). Measuring their internal rotation can thus place strong constraints on the efficiency of AM transport during this phase. This is what we set out to do in this paper. In Sect. \ref{sect_analysis}, we present two young subgiants observed with \kepler, for which we have been able to measure the rotational splittings of mixed modes. A seismic modeling of these targets is described in Sect. \ref{sect_model}. In Sect. \ref{sect_inversions}, we perform inversions of the rotation profiles of these stars. We then interpret our results to estimate the efficiency of the transport of AM during the subgiant phase in Sect. \ref{sect_transport}. Sect. \ref{sect_concl} is dedicated to conclusions.


\section{Signature of rotation in the oscillation spectra of two \kepler\ subgiants \label{sect_analysis}}

\subsection{Why measuring rotation in young subgiants is challenging}

For several reasons, young subgiants are less suited to rotation inversions than their more evolved counterparts. 

First, their oscillation spectra contain less g-dominated mixed modes. Owing to their lower core density, their \vaisala\ frequency is lower, and so are the frequencies of their gravity modes. The oscillation spectra of subgiants  after the main sequence turnoff result from the coupling of p modes with the lowest-order gravity modes ($n_{\rm g}=1,2,3,...$, where $n_{\rm g}$ is to the number of nodes of the mode eigenfunctions in the g-mode cavity). The frequency separation between consecutive-order g modes of low $n_{\rm g}$ is large and the oscillation spectra of subgiants thus contain only a few g-dominated modes. For this reason, we expect to measure the core rotation of subgiants less precisely than for red giants, which contain tens of g-dominated mixed modes.

Secondly, subgiants are hotter than red giants. It is now established from both observational and theoretical point of view that the damping of the oscillation modes increases very steeply with the effective temperature of the star (\citealt{appourchaux12}, \citealt{belkacem12}). Consequently, the oscillation modes of subgiants have shorter lifetimes, and thus larger line widths in the oscillation spectra. This makes the measurement of the mode splittings more complicated and statistical tests are needed to establish the significance of the rotational splittings.

Finally, the oscillation modes excited in subgiants have frequencies well above the Nyquist frequency of the long-cadence data of \kepler\ (278 $\mu$Hz) and short-cadence data (integration time of 58.84876 s) are thus required. This drastically limits the number of targets available in the \kepler\ catalog compared to the case of more evolved red giants.

\subsection{Selection of targets}

Among the subgiants observed with the \kepler\ satellite in short cadence, we searched for stars that are less evolved than the sample studied by D14 ($\log g > 3.85$) and for which the rotational splittings of mixed modes could be extracted with a high-enough significance level for rotation inversions. For this purpose, we selected stars among the catalog of \cite{chaplin14} with clear detection of mixed modes, duration of observations over one year (to ensure a good enough frequency resolution to measure individual mode splittings) and $\log g > 3.85$. To identify interesting targets among this sample, we performed a preliminary measurement of the rotational splittings for nonradial modes using a maximum likelihood estimation technique and performing statistical tests to determine whether or not the rotational splittings of the modes are significant. This step follows directly the method described in Sect. 3.1 of  \cite{deheuvels15}. We thus identified two subgiants that provided enough significant rotational splittings to perform inversions (KIC5955122, and KIC8524425). The known properties of these targets are described below. We note that applying more advanced statistical studies to all subgiants in the  \cite{chaplin14} catalog might lead to identify more targets of the same type.

\begin{figure*}
\begin{center}
\includegraphics[width=9cm]{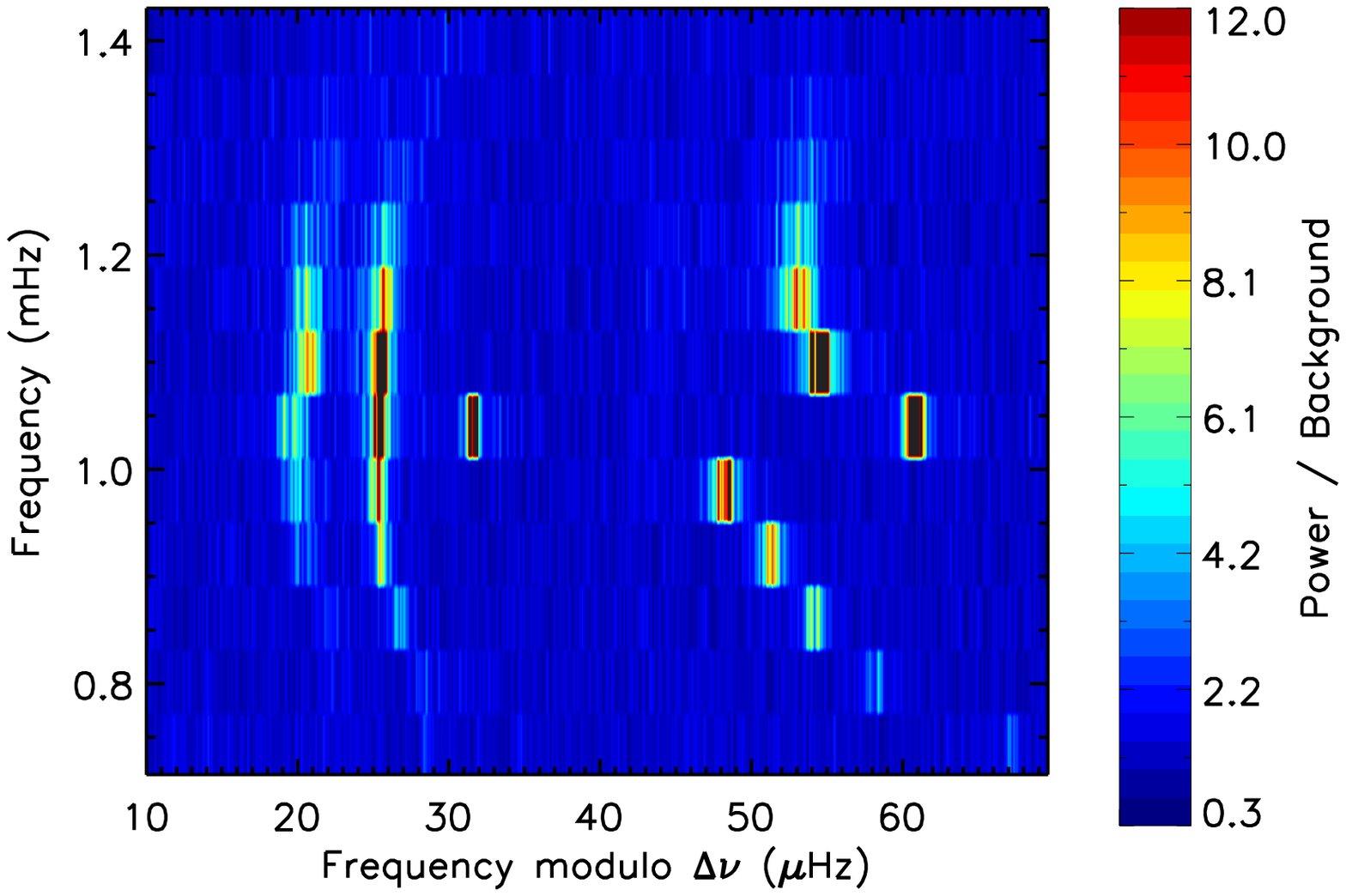}
\includegraphics[width=9cm]{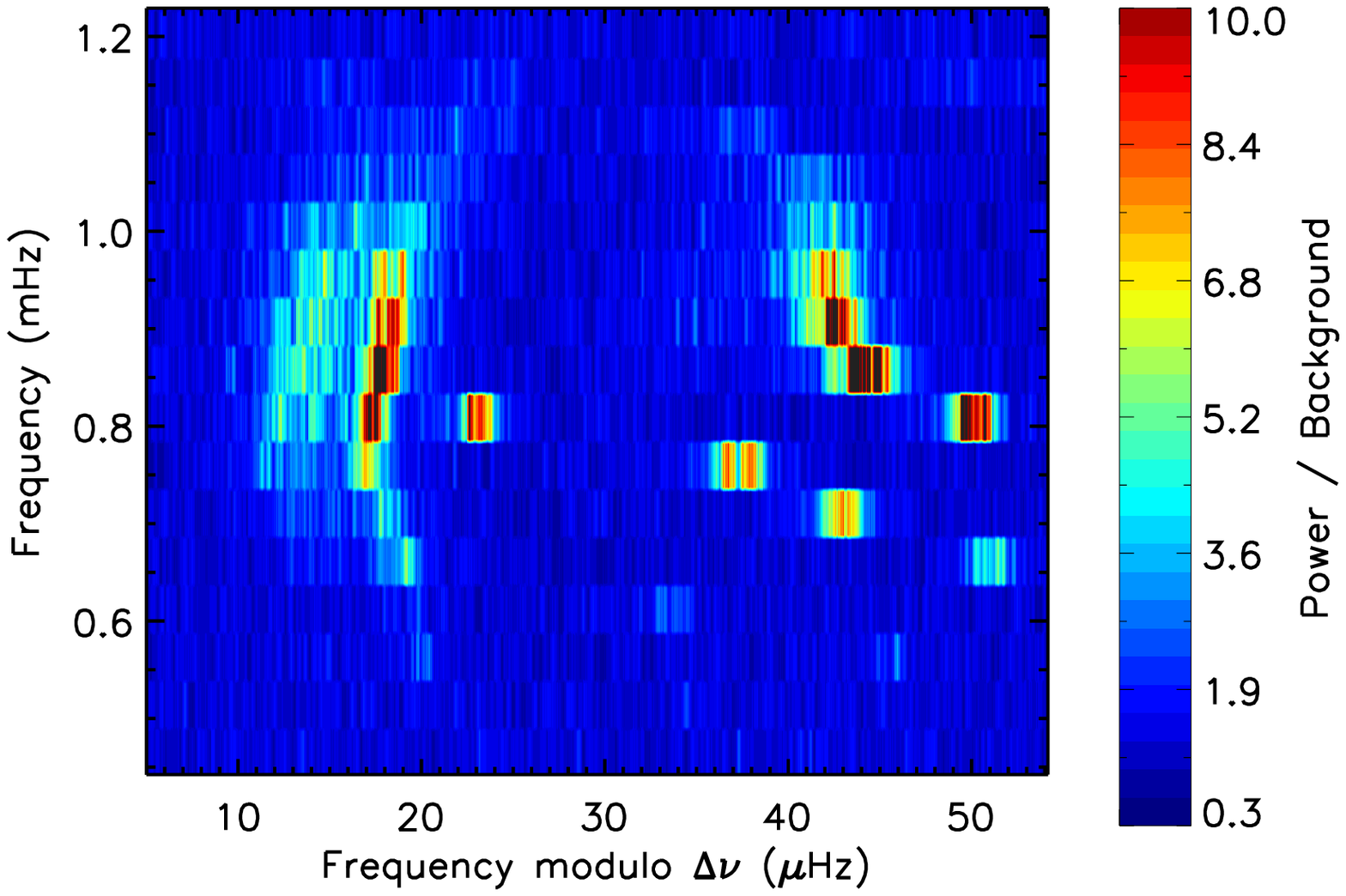}
\end{center}
\caption{Echelle diagrams of the power spectra of KIC8524425 (left panel) and KIC5955122 (right panel).
\label{fig_echelle_obs}}
\end{figure*}

\subsubsection{KIC5955122 \label{sect_595}}

KIC5955122 is an F9 star observed with the \kepler\ satellite in short cadence during quarters Q1 and then Q5 through Q17 (1181 days overall). Spectroscopic observations were obtained for this star by \cite{bruntt12}, who found $T_{\rm eff} = 5865\pm70$ K, $[$Fe/H$]=-0.17\pm0.06$ dex, $\log g = 3.88\pm0.08$ and $v\,\sin i = 6.5$ km.s$^{-1}$. The \kepler\ light curve of KIC5955122 shows clear signatures of spot modulations and the star is thus magnetically active. \cite{bonanno14} analyzed the light curve of the star and found that the lifetimes of spots range from 16.4 days to 27.7 days. They attributed these variations to latitudinal differential rotation at the surface of the star, which they estimated to 0.25 rad.d$^{-1}$ using spot modeling. \cite{garcia14} found a rotation period of $19.13\pm2.41$ days for this star using the spot-modulation of the \kepler\ light curve, which is in agreement with the measurement by \cite{bonanno14}.

Using adaptative optics at Subaru, \cite{schonhut17} found that KIC5955122 has an M-type companion source with a separation of $3.60\pm0.06''$. The Gaia DR2 indicates that this companion has a similar parallax as KIC5955122 ($5.53\pm0.02$ mas for KIC5955122, $5.51\pm0.05$ mas for the companion, \citealt{gaia_dr2}), which means that the two stars likely form a wide binary system. To estimate the star's luminosity, we combined the Gaia parallax with the magnitude $m_{\rm G}=9.24$ in the Gaia passband. The bolometric correction and the effects of reddening are estimated following \cite{casagrande18}. Assuming a a reddening of $E(B-V)=0.03^{+0.02}_{-0.03}$ from the three-dimensional dust map of \cite{green18}, we obtained a luminosity estimate of $L/L_\odot = 5.3\pm0.4$ for the star.


\subsubsection{KIC8524425 \label{sect_852}}

KIC8524425 is a G2 target of the \kepler\ mission observed during quarter Q2 and then continuously during quarters Q5 through Q17. Using spectroscopy, \cite{bruntt12} found $T_{\rm eff} = 5620\pm70$ K, $[$Fe/H$]=0.14\pm0.06$ dex, $\log g = 4.03\pm0.08$ and $v\,\sin i = 2.3$ km.s$^{-1}$ for this star. Its Gaia parallax is $\pi=5.78\pm0.02$ mas (\citealt{gaia_dr2}), which means that its distance is $172.9\pm0.7$ pc. Using reddening from \cite{green18} ($E(B-V)=0.03^{+0.02}_{-0.03}$) and a bolometric correction from \cite{casagrande18}, we obtained $L/L_\odot = 3.1\pm0.2$ for this star. Spot modulations were found in the light curve of KIC8524425 by \cite{garcia14}. Using those, they could derive a surface rotation period of $42.44\pm3.44$ days for the star.

\subsection{Seismic analysis \label{sect_sismo}}

A seismic analysis of the two targets has already been performed by \cite{appourchaux12b} using the \kepler\ data available at that time (quarters Q5--Q7, corresponding to 275 days of data). They applied statistical tests to detect oscillation modes and extracted their frequencies from the oscillation spectra. The frequency resolution at that time was not sufficient to measure individual rotational splittings of the modes. We here performed a new analysis of the oscillation spectra of the two stars using the complete \kepler\ dataset with the aim to (i) increase the precision of the mode frequency estimates compared to \cite{appourchaux12} and (ii) to measure the rotational splittings of the mixed modes. The power spectra of the two stars are shown in the shape of \'echelle diagrams in Fig. \ref{fig_echelle_obs}. We followed the procedure that was described in \cite{deheuvels15}, which is only briefly recalled here.

\subsubsection{Fit of the background}

The light curves of the two stars were processed using the \kepler\ pipeline developed by \cite{jenkins10}. Corrections from outliers, occasional jumps, and drifts were applied following \cite{garcia11}. The power density spectra (PSD) were then obtained using the Lomb-Scargle periodogram (\citealt{lomb76}, \citealt{scargle82}). Prior to fitting individual oscillation modes, we fit the background of the PSD using a maximum-likelihood estimation (MLE) method. Our model of the background contains contributions from the granulation, the photon noise, and the oscillations. The contribution from the granulation was modeled by two Harvey profiles (\citealt{harvey85}) with distinct characteristic timescales, as advocated by \cite{karoff13} and \cite{kallinger14}. We also added a white noise component to account for photon noise. The contribution from the oscillations was modeled as a Gaussian function. The background model was used to build the prior probability law for the background level when extracting the mode parameters. The central frequency of the Gaussian component gives an estimate of the frequency of maximum power of the oscillations $\nu_{\rm max}$. We obtained $\nu_{\rm max} = 860\pm12 \,\mu$Hz for KIC5955122, and $\nu_{\rm max} = 1091\pm10 \,\mu$Hz for KIC8524425. These values were used when applying seismic scaling relations (see Sect. \ref{sect_scaling}).

\subsubsection{Modeling of the oscillation spectrum}
\label{sect:modspect}

Each oscillation mode was modeled as a Lorentzian function in the power spectrum, characterized by its central frequency $\nu_{n,l,m}$, its height $H$, and its 
full width at half maximum 
$\Gamma$. The height ratios of the $m$-components within multiplets (modes of same radial order $n$ and degree $l$) depend on the stellar inclination angle $i$ and their expressions were derived using the relations given by \cite{gizon03}. The different $m$-components of the rotational multiplets  were assumed to have a common line width $\Gamma$. 

Owing to the slow rotation of subgiants and red giants, the rotational splittings can be expressed as a first-order perturbation to the mode frequencies. Previous rotation inversions performed for subgiants and red giants have all assumed that the rotation profile is spherically symmetric (e.g., \citealt{deheuvels12}, \citealt{dimauro16}). However, the modulations in the light curve of KIC5955122 due to spots suggest that its convective envelope shows latitudinal differential rotation (see Sect. \ref{sect_595}). In this case, the symmetric rotational splittings defined as $S_{n,l,m} = (\nu_{n,l,m}-\nu_{n,l,-m})/(2m)$ are expected to vary with the azimuthal order $m$. This needs to be taken into account in order to reliably extract the rotational splittings of quadrupole multiplets.

We expressed the rotational splittings as a sum of orthogonal polynomials using the Clebsch-Gordon $a$-coefficient decomposition, as was done in helioseismology (\citealt{ritzwoller91}) and more recently for main-sequence solar-like pulsators (\citealt{benomar18}). We then have 
\begin{equation}
\nu_{n,l,m} = \nu_{n,l} + \sum_{i=1}^{2l+1} a_i^{(l)}(n) \mathcal{P}_i^{(l)}(m),
\label{eq_ritzwoller}
\end{equation}
where the $\nu_{n,l}$ are the mode frequencies of the nonrotating star and the $\mathcal{P}_i^{(l)}(m)$ are orthogonal polynomials of degree $i$ that are explicitly given by \cite{pijpers97}. Only the $a_i$ coefficients with odd $i$ need to be considered because of the symmetry properties of the splittings.
Having access to modes with degrees $l\leqslant2$, we can only measure $a_1$ and $a_3$ coefficients here. The symmetric rotational splittings are then given by
\begin{align}
S_{n,1,1} & = a_1^{(1)}(n)  \\
S_{n,2,m} & = a_1^{(2)}(n) +  \frac{1}{3} (5m^2-17) a_3^{(2)}(n).
\end{align}
The advantage of this decomposition is that the $a_1$ coefficients probe the internal rotation, while the $a_3$ coefficients measure the latitudinal differential rotation.

The quality of our data is not sufficient to measure individual values of $S_{n,2,m}$ for the components of each quadrupole multiplets. We therefore had to make simplifications. We tested two different models: 
\begin{enumerate}
\item \textit{Model $M_1$:} It is well known that most $l=2$ modes that are detected in subgiants and red giants are p-dominated because the coupling between the p- and g-cavities is weak. Our modeling of both subgiants in Sect. \ref{sect_model} confirms that the contribution from the core to the kinetic energy of the detected quadrupole modes is indeed very small. In these conditions, it is fair to assume that quadrupole modes probe only the rotation in the p-mode cavity. As for main-sequence solar-like pulsators, the rotational kernels of quadrupole modes with different radial orders have roughly the same sensitivity to the rotation profile. We thus chose to assume a common value of $a_1^{(2)}$ and $a_3^{(2)}$ for all quadrupole modes in order to make the fit more robust.\\
\item \textit{Model $M_2$:} We measured individual rotational splittings for quadrupole modes, neglecting the effects of latitudinal rotation (that is, setting $a_3^{(2)}=0$). 
\end{enumerate}

We note that we have here neglected the effects of near-degeneracies in the mode frequencies, which can in some cases create asymmetries in the rotational multiplets (\citealt{deheuvels17}). In practice, asymmetries arise when the frequency separation between consecutive mixed modes is comparable to the rotational splitting. This is not the case for our stars here (see Sect. \ref{sect_seismic_model}), so we could safely neglect near-degeneracy effects. 

\subsubsection{Extraction of mode parameters}

We then proceeded to fit the individual oscillation modes with a Bayesian approach \citep[e.g.,][]{benomar09a}. We estimated the parameters $\boldsymbol{\theta}$ for a spectrum model $M_i$ (described in the Sect.~\ref{sect:modspect}) given a data set $D$ and prior information $I$ through posterior probability using Bayes theorem:
\begin{equation}
 p(\boldsymbol{\theta}|D,M_i,I)=\frac{p(\boldsymbol{\theta}|M_i,I)p(D|\boldsymbol{\theta},M_i,I)}{p(D|M_i,I)},
\end{equation}
where $p(\boldsymbol{\theta}|M_i,I)$ is a prior probability law,
$p(D|\boldsymbol{\theta},M_i,I)$ is the likelihood and $p(D|M_i,I)$ the evidence of model $M_i$ knowing the observation $D$, defined as the marginalization over the whole parameter space of the product of the likelihood and the prior. We assumed that the observed spectrum follows the distribution of a $\chi^2$ with 2 degrees of freedom.

We performed semi-local fits. We defined small windows of 16~$\mu$Hz in the vicinity of each oscillation modes (the size of the windows was chosen such that the local background level can be efficiently estimated). When two or three modes are too close to be isolated, they have been fit together in a broader window. Thus, $l=0$ and $l=2$ modes have been fit together, and sometimes, a close-by $l=1$ mode has also been included. For each box, we assumed a constant background $B$. All the windows have been simultaneously fit with a common value for the inclination angle $i$. When fitting model $M_1$, we used a common value of $a_1^{(2)}$ and $a_3^{(2)}$ for all modes, as mentioned in Sect. \ref{sect:modspect}.

The prior probability laws for the different free parameters we used are very similar to the ones we chose in \citet{deheuvels15}. For the mode frequencies $\nu_{n,l}$, we used a uniform prior spanning an interval of 3 $\mu$Hz. For the mode heights $H$, widths $\Gamma$ and for the background level $B$, we used uniform priors for the parameters $\ln \Gamma$, $\ln(\pi H\Gamma/2)$ and $\ln B$. This is the same as using Jeffreys prior for $H$, $\Gamma$, and $B$. We assumed an isotropic prior for the rotation axis orientation, and thus considered a uniform prior for $\cos i$ over [0,1]. For the rotational splittings, we used a uniform prior for $a_1$ over [0,1$\mu$Hz]. For model $M_1$, where splitting asymmetries are taken into account, we also used a uniform prior for the ratio $a_3^{(2)}/a_1^{(2)}$ over [-0.15,0.2]. This is a very broad prior: beyond these limits $m=\pm 1$ components can become degenerated with $m=\pm 2$ or $m=0$ components.

Posterior distribution is sampled with a Markov Chain Monte Carlo. Our code is based on Metropolis-Hastings algorithm and is very similar to the one developed by \citet{benomar09a}. We use parallel tempering with 20 parallel chains to avoid the sampling being stuck in local maxima.

Using model $M_1$, we obtained rather loose constraints on the $a_3$ coefficients. We found $a_3 = 16^{+23}_{-25}$~nHz for KIC5955122, and $a_3 = 15^{+20}_{-31}$~nHz for KIC8524425. We show in Appendix \ref{app_a3} that our measurement of the $a_3$ coefficient for KIC5955122 is compatible with the photometric measurement of the latitudinal differential rotation obtained by \cite{bonanno14}. For both stars, our measurements of $a_3$ are compatible with zero, which means that we can safely neglect the effects of latitudinal differential rotation, as was done for model $M_2$. We thus retain only the results obtained with this model in the following sections. In this context, we no longer need to use the Clebsch-Gordon decomposition for the rotational splittings. We further denote the rotational splittings of the modes as $\delta\nu_{\rm s}$.

\begin{figure}
\begin{center}
\includegraphics[width=8cm]{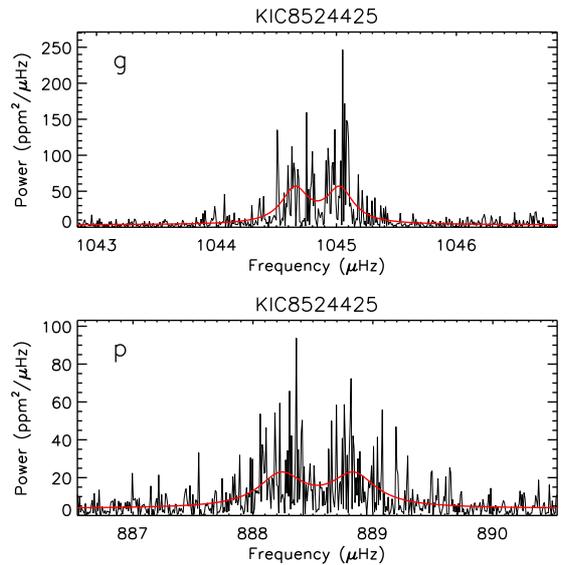}
\end{center}
\caption{Rotational multiplets of two dipolar mixed modes of KIC8524425. The top panel shows a g-dominated mode and the bottom panel, a p-dominated mode (see Sect. \ref{sect_zeta}). The red curve indicates the best fit to the data (see Sect. \ref{sect_sismo}). 
\label{fig_splittings_obs}}
\end{figure}

Both stars were found to have a relatively high inclination angle ($i=68^\circ$$^{+6.1}_{-4.3}$ for KIC5955122, and $i=78.1^\circ$$^{+7.9}_{-7.9}$ for KIC8524425). This makes these targets favorable to measure the rotation splittings of dipole mode because the $m=0$ components have a low visibility. Tables \ref{tab_852} and \ref{tab_595} in Appendix \ref{app_seismic_params} give the extracted mode frequencies for the two stars. To ensure robust results, we retained only the rotational splitting estimates for which the posterior distribution has a negligible amplitude around zero. For each parameter, we provide the median of the marginalized posterior distribution with errors that correspond to the 68\% probability interval. For illustration, Fig. \ref{fig_splittings_obs} shows two dipolar mixed modes of KIC8524425 that were found to be significantly split by rotation. The extracted mode frequencies for both stars are in good agreement with the recent estimates of \cite{li20_analysis}, who followed a similar approach.

\section{Seismic modeling \label{sect_model}}

\subsection{Seismic scaling relations \label{sect_scaling} }

Preliminary hints of the masses, radii, and surface gravities of the two stars can be derived from seismic scaling relations using measurements of $\nu_{\rm max}$ and the asymptotic large separation of p modes $\Delta\nu$. Estimates of $\nu_{\rm max}$ have already been obtained in Sect. \ref{sect_sismo}. To determine $\Delta\nu$, we fit a second-order asymptotic expression to the frequencies of the observed radial modes that were obtained in Sect. \ref{sect_sismo}. The asymptotic expression of the radial modes was taken following \cite{mosser13}, 
\begin{equation}
\nu_{n,l=0} = \left[ n+\varepsilon_{\rm p}+\frac{\alpha}{2}(n-n_{\max})^2 \right] \Delta\nu,
\end{equation}
where $n_{\rm max}$ corresponds to the radial order that satisfies $\nu_{n,l=0} \approx \nu_{\rm max}$. An estimate of the asymptotic large separation $\Delta\nu_{\rm as}$ was then obtained as $\Delta\nu_{\rm as} = \Delta\nu\left( 1+n_{\rm max}\alpha/2 \right)$ using Eq. 11 of \cite{mosser13}. We thus obtained $\Delta\nu_{\rm as} = 51.91\pm0.02 \,\mu$Hz for KIC5955122 and $\Delta\nu_{\rm as} = 61.72\pm0.02 \,\mu$Hz for KIC8524425. By combining our estimates of $\Delta\nu_{\rm as}$ and $\nu_{\rm max}$ with the measurements of $T_{\rm eff}$ given in Sect. \ref{sect_595} and \ref{sect_852}, we applied seismic scaling relations to the two stars. The results are given in Table \ref{tab_scaling}.  These rough estimates were refined when performing a full seismic modeling of the two targets in Sect. \ref{sect_seismic_model}.

\begin{table}
  \begin{center}
  \caption{Global stellar parameters obtained from seismic scaling relations. \label{tab_scaling}}
  \vspace{0.2cm}
\begin{tabular}{l c c c}
\hline \hline
\T \B Starname & $M/M_\odot$ & $R/R_\odot$ & $\log g$ \\
\hline
\T KIC5955122 & $1.11\pm0.07$ & $2.00\pm0.05$ & $3.884\pm0.020$  \\
\B KIC8524425 & $1.07\pm0.05$ & $1.75\pm0.03$ & $3.978\pm0.015$ \\ 
\hline
\hline
\end{tabular}
\vspace{0.2cm}

\end{center}
\end{table}


\subsection{Seismic modeling using individual mode frequencies \label{sect_seismic_model}}

\begin{figure*}
\begin{center}
\includegraphics[width=9cm]{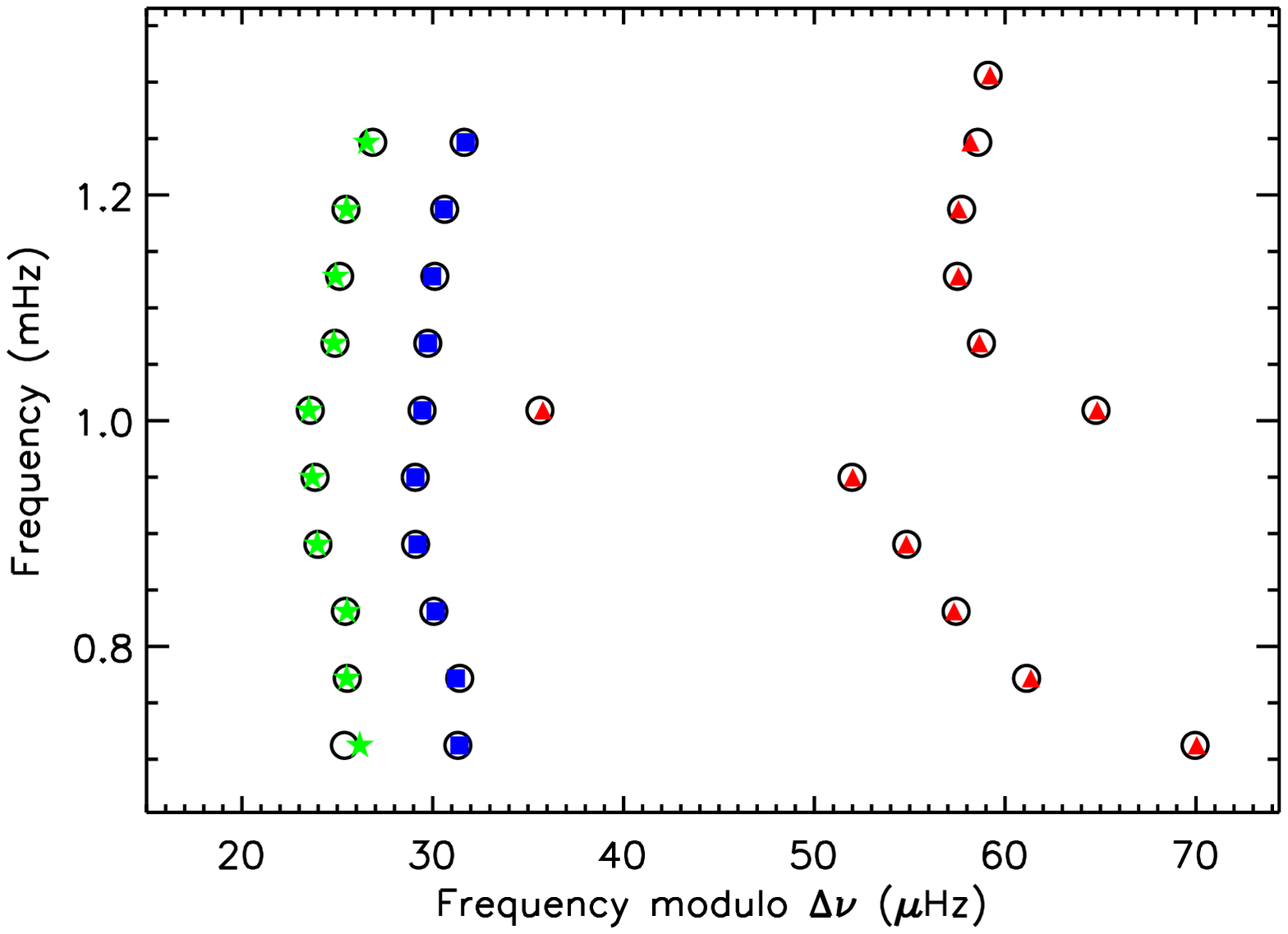}
\includegraphics[width=9cm]{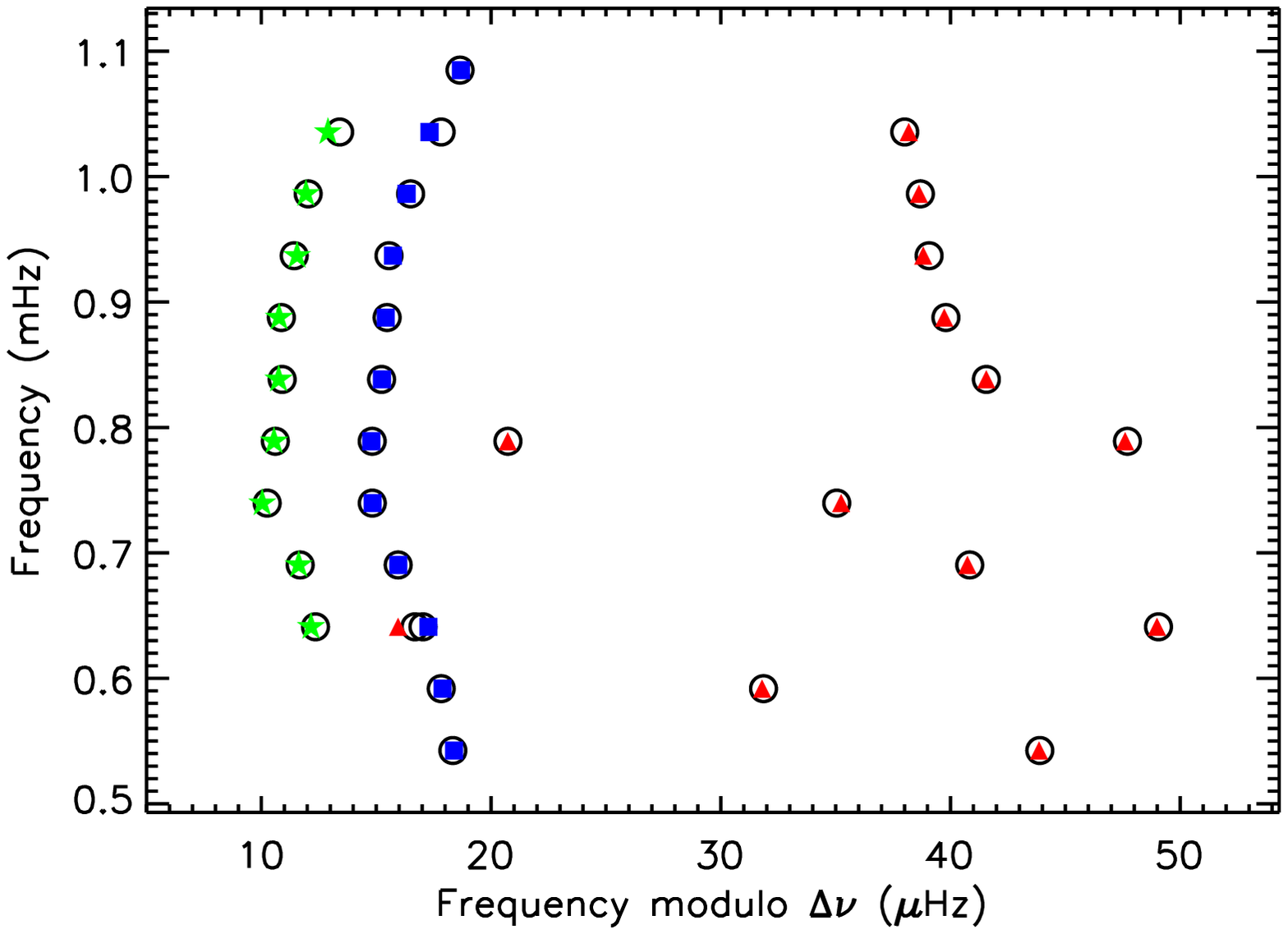}
\end{center}
\caption{Echelle diagrams of the best-fit model obtained for KIC8524425 (left panel) and KIC5955122 (right panel). The open circles correspond to observed mode frequencies and the filled symbols to the frequencies of the best-fit model (blue squares for $l=0$ modes, red triangles for $l=1$ modes, and green stars for $l=2$ modes).
\label{fig_echelle_mod}}
\end{figure*}

For red giants, the g modes that couple to the observed p modes have high radial orders $n_{\rm g}$, so that the trapping of the modes can be estimated directly from the observed oscillation spectra, using the mode frequencies and asymptotic expressions  (\citealt{goupil13}). For young subgiants such as KIC5955122 and KIC8524425, which have $n_{\rm g}$ of order unity, this is not possible and one has to compute a stellar model that matches reasonably well the observed mode frequencies. 

For this purpose, we have used the \mesa\ stellar evolution code (version 10108, \citealt{paxton15}). We used the OPAL 2005 equation of state (\citealt{rogers02}) and opacity tables. The nuclear reaction rates were computed using the NACRE compilation (\citealt{angulo99}), except for the $^{14}$N$(p,\gamma)^{15}$O reaction where we adopted the revised LUNA rate (\citealt{formicola04}). The atmosphere was described by Eddington's gray law. We assumed the solar mixture of heavy elements of \cite{asplund09}. Convection was described using the classical mixing length theory (\citealt{bohm58}) with an adjustable mixing length parameter $\alpha_{\rm MLT}$. Microscopic diffusion was included by solving the equations of \cite{burgers69} at each time step. We have computed models with or without convective overshooting. For both stars, we found that models computed with a significant amount of overshooting led to similar or worse fits to the observations compared to models that do not include overshooting. Indeed, the masses of the two stars indicate that they had either a small convective core or a radiative core during the main sequence. In this mass range, low amounts of core overshooting are expected (\citealt{deheuvels16}). The eigenfrequencies of the stellar models were calculated using the \textsc{ADIPLS} code (\citealt{adipls}). The mode frequencies were then corrected from near-surface effects using the parametrized correction advocated by \cite{ball14} (cubic term proportional to $\nu^3/\mathcal{I}$, where $\nu$ is the mode frequency and $\mathcal{I}$ is its inertia).

Modeling subgiants with mixed modes is notoriously difficult. The frequencies of g-dominated modes vary on timescales that are short compared to the evolution timescale, which is problematic for usual seismic modeling techniques. To model the two stars under study, we followed the method described by \cite{deheuvels11}. For a given set of input stellar parameters (initial helium and heavy element abundances, mixing-length parameter), the authors have shown that the frequencies of radial modes and the frequency of the most g-dominated mode can be used together to obtain very precise estimates of the stellar mass and age. For both stars, we have thus computed grids of models with varying $Y_0$, $(Z/X)_0$, and $\alpha_{\rm MLT}$. For each grid point, we performed an optimization yielding the best-fit mass and age, as mentioned above. A traditional grid-based approach, where mass and age are considered as free parameters, would have required tiny steps in mass and age (and therefore a tremendous number of models) in order to catch the best-fit models. 

\begin{figure}
\begin{center}
\includegraphics[width=9cm]{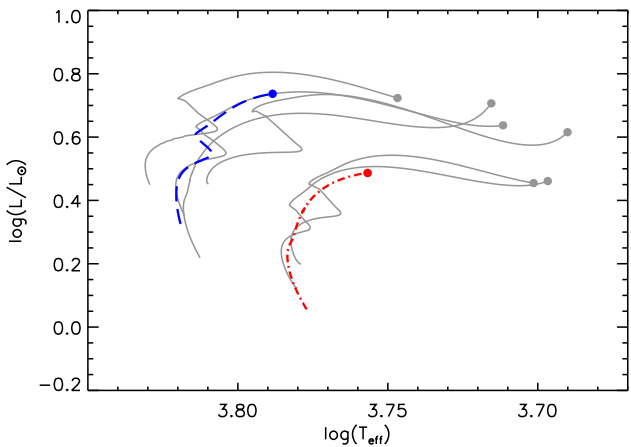}
\end{center}
\caption{Evolutionary tracks in the HR diagram of the optimal stellar models of KIC5955122 (blue long-dashed curve) and KIC8524425 (red dot-dashed curve). Also shown in gray are the evolutionary tracks of the stars whose rotation profiles were measured by D14.
\label{fig_hr_model}}
\end{figure}

\begin{table}
  \begin{center}
  \caption{Stellar parameters of the best-fit models of KIC5955122 and KIC8524425. \label{tab_model}}
  \vspace{0.2cm}
\begin{tabular}{l c c c}
\hline \hline
\T \B   & KIC5955122 & KIC8524425 \\
\hline
\T Mass ($M_\odot$) & $1.218$ & 1.113 \\
Age (Gyr) & $4.65$ & 7.58 \\
Radius ($R_\odot$) & $2.099$ & 1.797 \\
Luminosity ($L_\odot$) & 5.49 & 3.06 \\
$(Z/X)_0$ & $0.0143$ & 0.0266  \\
$Y_0$ & 0.257 & 0.277 \\
\B $\alpha_{\rm MLT}$ & $2.00$ & 1.86 \\
\hline
\T \B $\chi^2_{\rm red}$ & 2.0 & 5.1  \\
\hline
\hline
\end{tabular}
\vspace{0.2cm}
\end{center}
\end{table}

\begin{figure*}
\begin{center}
\includegraphics[width=9cm]{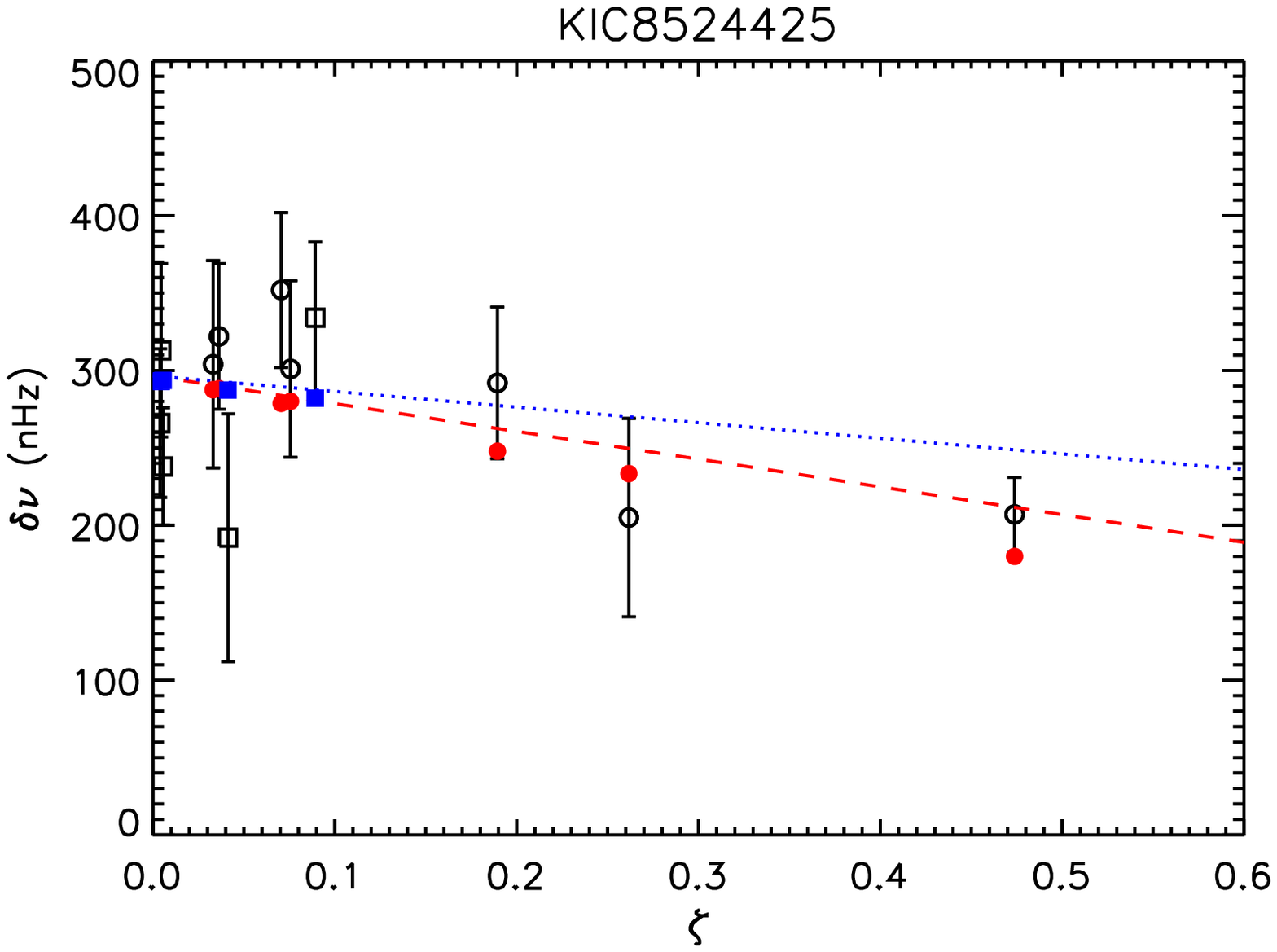}
\includegraphics[width=9cm]{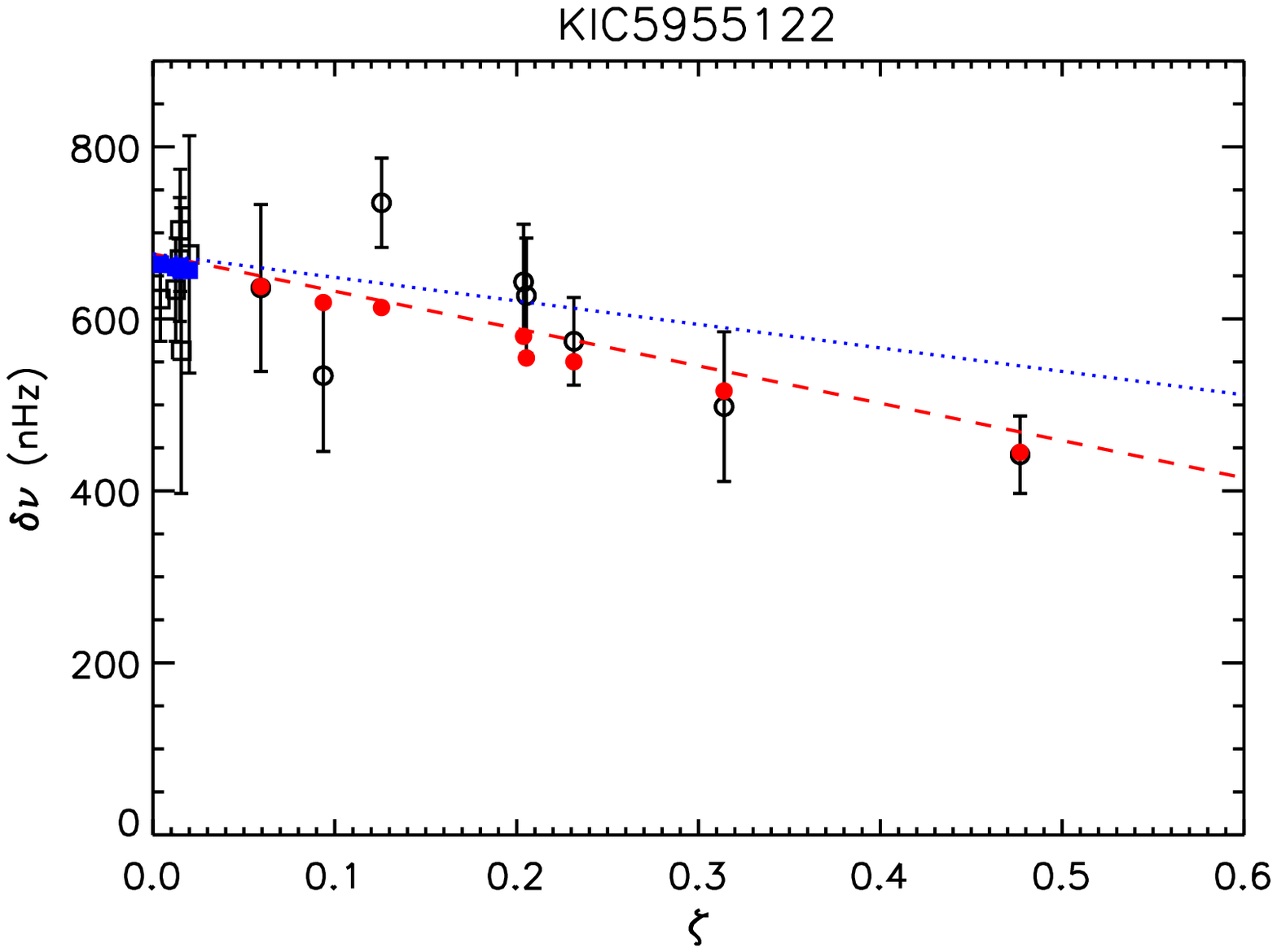}
\end{center}
\caption{Observed rotational splittings (open symbols) for modes of degrees $l=1$ (circles) and $l=2$ (squares) plotted as a function of the parameter $\zeta$ obtained from the best-fit stellar models ($\zeta$ measures the trapping of the modes, with $\zeta\rightarrow0$ for p-dominated and $\zeta\rightarrow1$ for g-dominated modes). The splittings obtained with the inverted values of $\omg$ and $\omp$ (see Sect. \ref{sect_inversions}) are shown as filled symbols. The dashed and dotted lines show the expected rotational splittings for $l=1$ and 2 modes, respectively, when fitting Eq. \ref{eq_g13} to the observations. 
\label{fig_zeta_split}}
\end{figure*}

For each model of the computed grids, we estimated the agreement with the observations by computing the $\chi^2$ function defined as
\begin{equation}
\chi^2 = \sum_{i=1}^N \frac{(\mathcal{O}_i^{\rm mod} - \mathcal{O}_i^{\rm obs})^2}{\sigma_i^2},
\end{equation}
where $\mathcal{O}_i^{\rm obs}$, $i=1,\hdots,N$ are the $N$ observables that were used to constrain the models, $\sigma_i$ correspond to the associated uncertainties, and $\mathcal{O}_i^{\rm mod}$ are the corresponding values in the computed stellar models. In this case, the observables include to the global surface parameters ($T_{\rm eff}$, $L$, $\log g$, and $(Z/X)_{\rm surf}$) and the frequencies of the $l=0,1,$ and 2 modes given in Tables \ref{tab_852} and \ref{tab_595}. The stellar parameters of the best-fit models are given in Table \ref{tab_model}. As can be seen in Fig. \ref{fig_echelle_mod} the mode eigenfrequencies of the optimal models (corrected from near-surface effects) show a very good visual agreement with the observations. The reduced $\chi^2$ of these best-fit models are $\chi^2_{\rm red}=2.0$ for KIC5955122, and $\chi^2_{\rm red}=5.1$ for KIC5955122. For these two models, the $\chi^2$ value is dominated by the contribution of a few mixed modes of degree $l=1$ or 2. 

Seismic modelings of the two subgiants have also recently been performed by \cite{li20_modeling}, who worked with precomputed grids of stellar models, using an interpolation of the mixed mode frequencies with age. Our best-fit models are significantly more massive than those of \cite{li20_modeling}, who found $M = 0.97\pm0.04\,M_\odot$ for KIC8524425 and $M = 1.12\pm0.05\,M_\odot$ for KIC5955122. However, we notice that the agreement with the observed mode frequencies, and in particular with the mixed dipole modes, is much better in the present study than for the best-fit models of \cite{li20_modeling} (comparing Fig. \ref{fig_echelle_mod} to the \'echelle diagrams in Appendix B of \citealt{li20_modeling}). The authors attribute these large differences between models and observations to an imprecise modeling of the core. A more detailed analysis would be required to identify the source of this disagreement, but this could be related to the fact that they only considered models with core overshooting, while our best-fit models for both stars were found for negligible core overshooting. Microscopic diffusion, which was neglected in the study of \cite{li20_modeling}, could also account for part of this disagreement.

In practice, it has been shown that the results of rotation inversions does not depend critically on the choice of a best-fit model, provided the mode frequencies are fit reasonably well (\citealt{deheuvels12}). When inferring the internal rotation profile in Sect. \ref{sect_inversions}, we have also checked that our conclusions are unchanged if we choose other models with comparable $\chi^2$ values from the grid as reference models for our inversions. Fig. \ref{fig_hr_model} shows the evolutionary tracks of the optimal models for the two stars in an HR diagram, compared to those of the stars studied by D14. It is readily seen that KIC8524425 and KIC5955122 are indeed less evolved than the D14 sample.

Our optimal models for KIC5955122 all have a convective core during the main sequence, while the core of KIC8524425 was radiative. This is important to stress because stars that have a convective core during the main sequence experience a rapid core contraction outside of thermal equilibrium as the star adjusts to the stopping of nuclear reactions in the core. Without redistribution of AM, this contraction induces a sharp increase in the core rotation rate (\citealt{eggenberger19}).

Finally, using our best-fit models for the two stars, we could estimate the minimum separation between consecutive mixed modes, in order to test whether asymmetries in rotational multiplets due to near-degeneracy effects are expected (\citealt{deheuvels17}). For KIC8524425, we found minimum separations of $30.7\,\mu$Hz for $l=1$ modes and $10.7\,\mu$Hz for $l=2$ modes. For KIC5955122, we obtained $25.1\,\mu$Hz for $l=1$ modes and $5.5\,\mu$Hz for $l=2$ modes. These separations are much larger than the measured rotational splittings, so we confirm that neglecting near-degeneracy effects, as was done in Sect. \ref{sect_sismo}, is legitimate here.

\subsection{Mode trapping vs rotational splittings \label{sect_zeta}}

Using the optimal stellar models for the two stars, we could estimate the trapping of the observed modes. This is traditionally done using the $\zeta$ parameter, which measures the contribution of the g-mode cavity to the mode energy
\begin{equation}
\zeta = \frac{\int_{r_{\rm a}}^{r_{\rm b}} \rho r^2 (\xi_r^2 + L^2 \xi_h^2) \,\hbox{d}r}{\int_0^R \rho r^2 (\xi_r^2 + L^2 \xi_h^2) \,\hbox{d}r},
\end{equation}
where $\xi_r$ and $\xi_h$ are the radial and horizontal displacements of the mode, $r_{\rm a}$ and $r_{\rm b}$ are the turning points bounding the g-mode cavity, and $L^2 = l(l+1)$. We calculated $\zeta$ for all modes and we confirmed that the detected quadrupole modes are indeed p-dominated (we found $\zeta < 0.09$ for KIC8524425 and $\zeta < 0.02$ for KIC5955122). We note that for more evolved targets, estimates of $\zeta$ can be obtained directly from the mode frequencies using an asymptotic analysis (\citealt{goupil13}). However, in our case the low number of nodes of the modes in the g-mode cavity ($n_{\rm g}$ is order unity) prevents the use of an asymptotic treatment and $\zeta$ has to be calculated from a stellar model.

\cite{goupil13} showed that the splittings of dipole modes are expected to follow a linear trend with $\zeta$. Their expression can easily be extended to modes of degree $l$, giving
\begin{equation}
\delta\nu_{\rm s} \approx \frac{L^2-1}{L^2} \zeta \frac{\omg}{2\pi} + (1-\zeta) \frac{\omp}{2\pi},
\label{eq_g13}
\end{equation}
where $\omp$ and $\omg$ are average rotation rates in the p- and g-mode cavities. The observed rotational splittings are plotted as a function of $\zeta$ in Fig. \ref{fig_zeta_split}. 

Dipole mode splittings are indeed found to vary roughly linearly with $\zeta$. One striking observation is that in contrast with all previously studied subgiants and red giants, the splittings of p-dominated modes ($\zeta\rightarrow0$) are larger than those of g-dominated modes ($\zeta\rightarrow1$). This is already an indication that the core is not rotating much faster than the envelope in these two stars. Preliminary estimates of the core and envelope rotation rates were obtained by fitting Eq. \ref{eq_g13} to the observed splittings. We obtained $\omg/(2\pi) = 224\pm107$~nHz and $\omp/(2\pi) = 288\pm18$~nHz for KIC8524425, and $\omg/(2\pi) = 535\pm272$~nHz and $\omp/(2\pi) = 656\pm34$~nHz for KIC5955122. This suggests that both stars are compatible with a solid-body rotation profile. These first estimates were then refined using full seismic inversions in Sect. \ref{sect_inversions}.


\section{Rotation inversions \label{sect_inversions}}

We then used the rotational kernels of the best-fit models to measure the internal rotation of the two stars. For this purpose, the optimally localized averages (OLA) inversion technique is particularly well suited. For a given location $r_0$ inside the star, it consists in building an averaging kernel $\mathcal{K}_{\rm av}(r_0;r) = \sum_{i=1}^M c_i(r_0) K_i(r)$, where the functions $K_i(r)$, $i=1,M$ correspond to the rotational kernels of the modes whose splittings have been measured (for modes of degrees $l=1$ and 2), and the coefficients $c_i$ are optimized so that $\mathcal{K}_{\rm av}(r_0;r)$ is as localized as possible around $r_0$. Different approaches have been proposed, whereby the averaging kernels are built to approximate at best a Dirac function (multiplicative optimally localized averages, or MOLA method) or another well chosen target function, such as a Gaussian function (subtractive optimally localized averages, or SOLA method). 

\subsection{Envelope rotation rate \label{sect_env_rot}}

Previous rotation inversions for subgiants and red giants have failed to build averaging kernels that resemble Dirac or Gaussian functions within the envelope. This is because the eigenfunctions of the observed modes in the p-mode cavity are too similar to one another to obtain localized information on the rotation profile in this region. The situation is in fact quite similar to that of main sequence solar-like pulsators, for which only an average rotation of the p-mode cavity can be obtained (e.g., \citealt{benomar15}). Thus, the MOLA and SOLA inversion methods are not directly suited to measure the envelope rotation.

To obtain an average rotation in the p-mode cavity, one needs to cancel at best the contribution of the g-mode cavity to the envelope averaging kernel. We thus propose a slight modification to the MOLA method. Instead of imposing a particular shape to the averaging kernel, we minimize the integral of $\mathcal{K}(r)^2$ below a radius $r_{\rm c}$ that is chosen within the evanescent region between the p- and g-mode cavities. We thus minimize the functional
\begin{equation}
S(\mathcal{K}) = \int_0^{r_{\rm c}} \mathcal{K}(r)^2 \,\hbox{d}r + \lambda \sum_{i=1}^M c_i^2\sigma_i^2,
\end{equation}
where the second term corresponds to a regularization term. It limits the error magnification that can arise when the optimal coefficients $c_i$ take on large values of opposite signs. The coefficient $\lambda$ is a trade-off parameter. We performed tests of this inversion method using artificial input rotation profiles, from which we found that $\lambda \sim 10^{-2}$ provides a good compromise between the minimization of  $\int_0^{r_{\rm c}} \mathcal{K}(r)^2 \,\hbox{d}r$ and error magnification. 

\begin{figure}
\begin{center}
\includegraphics[width=9cm]{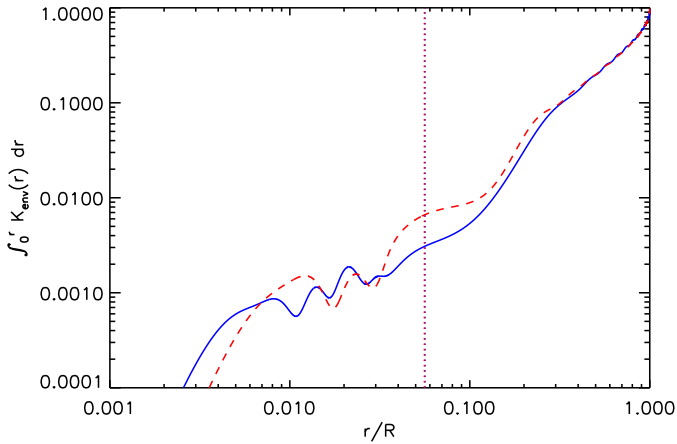}
\end{center}
\caption{Cumulative integral of the envelope averaging kernels obtained with the modified MOLA method (see text) for KIC8524425 (red dashed line) and KIC5955122 (solid blue line). The vertical dotted lines indicate the upper turning point of the g-mode cavity.
\label{fig_kernel_env}}
\end{figure}

\begin{figure}
\begin{center}
\includegraphics[width=9cm]{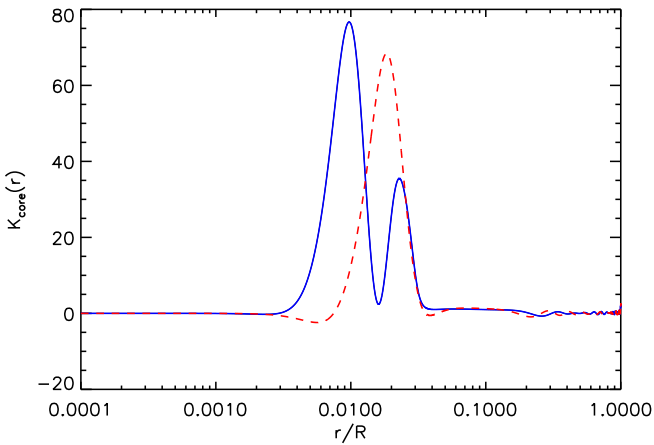}
\end{center}
\caption{Core averaging kernels obtained with the modified MOLA method (see text) for KIC8524425 (red dashed line) and KIC5955122 (solid blue line).
\label{fig_kernel_core}}
\end{figure}

Fig. \ref{fig_kernel_env} shows the cumulative integral of the envelope averaging kernels obtained by minimizing the function $S(\mathcal{K})$ for KIC8524425 and KIC5955122. As can be seen, the core contribution is below 1\% in both cases. We found that the $l=2$ rotational splittings are not crucial to obtain measurements of the average envelope rotation rate. Envelope averaging kernels with similar shapes as those shown in Fig. \ref{fig_kernel_env} can be obtained using $l=1$ modes only. However, the rotational splittings of $l=2$ modes improve the precision on the measurements of the average envelope rotation by a factor of 1.5 (KIC8524425) to 2 (KIC5955122). We were thus able to derive average rotation rates in the p-mode cavity of both stars. We obtained $\langle\Omega\rangle_{\rm p}/(2\pi) = 298\pm 20$~nHz for KIC8524425 and  $\langle\Omega\rangle_{\rm p}/(2\pi) = 675\pm 27$~nHz for KIC5955122. This corresponds to periods of rotation of $\langle P\rangle_{\rm p} = 38.8\pm2.6$ days for KIC8524425 and $\langle P\rangle_{\rm p} = 17.1\pm0.7$ days for KIC5955122. 


These rotation periods $\langle P\rangle_{\rm p}$ agree within 1-$\sigma$ errors with the surface rotation periods that were obtained by \cite{garcia14} from the photometric signature of stellar activity ($19.13\pm2.41$ days for KIC5955122 and $42.44\pm3.44$ days for KIC8524425). A surface rotation period can also be derived from the spectroscopic estimates of $v\sin i$ obtained by \cite{bruntt12} for the two stars. By combining the $v\sin i$ measurements with the inclination angles that we obtained from our analyses of the oscillation spectra (Sect. \ref{sect_sismo}) and the stars' radii from our modeling (Sect. \ref{sect_seismic_model}), we obtained rotation periods of 15.2 days for KIC5955122 and 38.7 days for KIC8524425 (no error bars are given for the $v\sin i$ in \citealt{bruntt12}). These values also agree well with our seismic measurement of the average rotation in the p-mode cavity.

As can be seen in Fig. \ref{fig_kernel_env}, seismic measurements mostly probe the rotation in the outer layers of the star but they are nonetheless also sensitive to the rotation in the bulk of the convective envelope and to a lesser extent in the radiative region below. The mean radius of sensitivity of the envelope averaging kernels can be computed as $x_{\rm env} = \int_0^R x\mathcal{K}_{\rm env}(r)\,\hbox{d}r$, where $x=r/R$ is the normalized radius. For both stars, one finds $x_{\rm env}\approx$0.76. The good agreement between the seismic measurement of the average rotation in the p-mode cavity and the surface rotation period suggests that there can only be a mild differential rotation within the envelope.

\subsection{Core rotation rate \label{sect_core_rot}}

\begin{table*}
\caption{Estimates of the mean rotation rates $\omg$, $\omp$, and the ratio between these quantities obtained from the coefficients of the $\delta\nu(\zeta)$ relation or rotation inversions.}
\begin{center}
\begin{tabular}{l c c c c c c c c}
\hline \hline
\T\B Star & \multicolumn{2}{c}{$\omg/(2\pi)$ (nHz)} & & \multicolumn{2}{c}{$\omp/(2\pi)$ (nHz)} & & \multicolumn{2}{c}{$\omg/\omp$} \\
\cline{2-3}
\cline{5-6}
\cline{8-9}
\T\B &  $\delta\nu(\zeta)$ & inversions & & $\delta\nu(\zeta)$ & OLA & & $\delta\nu(\zeta)$ & inversions \\
\hline
\T KIC8524425 & $224\pm107$ & $204\pm134$ & & $288\pm18$ & $298\pm20$ & &  $0.78\pm0.42$ &  $0.68\pm0.47$  \\
\B KIC5955122 & $535\pm272$ & $488\pm227$ & & $656\pm34$ & $675\pm27$ & &  $0.82\pm0.46$ &  $0.72\pm0.37$  \\
\hline
\end{tabular}
\end{center}
\label{tab_rotation}
\end{table*}%

The rotation inversions performed so far on subgiants and red giants yielded very precise constraints on the core rotation rate (e.g., \citealt{deheuvels12}, \citealt{dimauro16}). This was achieved thanks to the detection of a large number of g-dominated modes. In our case, young subgiants have only few g-dominated modes, so it is also more difficult to build averaging kernels that efficiently suppress the contribution from the envelope. By testing the inversion method on simulated data, we found that building core averaging kernels with negligible contribution from the envelope requires to measure the rotational splittings of several individual $l=2$ modes. 

We applied the same inversion technique as in Sect. \ref{sect_env_rot} but this time we minimized the contribution from the envelope to the averaging kernel. The core averaging kernels $\mathcal{K}_{\rm core}$ thus obtained are shown in Fig. \ref{fig_kernel_core}. The contribution of the envelope is indeed efficiently canceled for both stars. Using these kernels, we obtained $\langle\Omega\rangle_{\rm g}/(2\pi) = 204\pm 134$~nHz for KIC8524425 and  $\langle\Omega\rangle_{\rm g}/(2\pi) = 488\pm 227$~nHz for KIC5955122. Contrary to rotation inversions performed on more evolved targets, the estimates of the core rotation rate for these two young subgiants are less precise than the measurement of the envelope rotation rate. 


We used the values of $\omg$ and $\omp$ obtained from seismic inversions to calculate the corresponding rotational splittings as 
\begin{equation}
\delta\nu_{\rm s} \approx \omg\int_0^{r_{\rm c}} K_{n,l}(r)\,\hbox{d}r + \omp\int_{r_{\rm c}}^R K_{n,l}(r)\,\hbox{d}r.
\label{eq_twozone}
\end{equation}
The results are overplotted in Fig. \ref{fig_zeta_split}. They match quite well the observed values of the $a_1$ coefficients, which brings further validation to our inversion results.



From the seismic measurements of $\omp$ and $\omg$, we could deduce estimates of the core-envelope contrast for the two stars. We obtained $\omg/\omp = 0.68\pm0.47$ for KIC8524425 and $\omg/\omp = 0.72\pm0.37$ for KIC56955122. This confirms that the two stars are indeed consistent with a solid-body rotation within the entire interior. 



\section{Interpretation for the transport of AM \label{sect_transport}}

\begin{figure}
\begin{center}
\includegraphics[width=9cm]{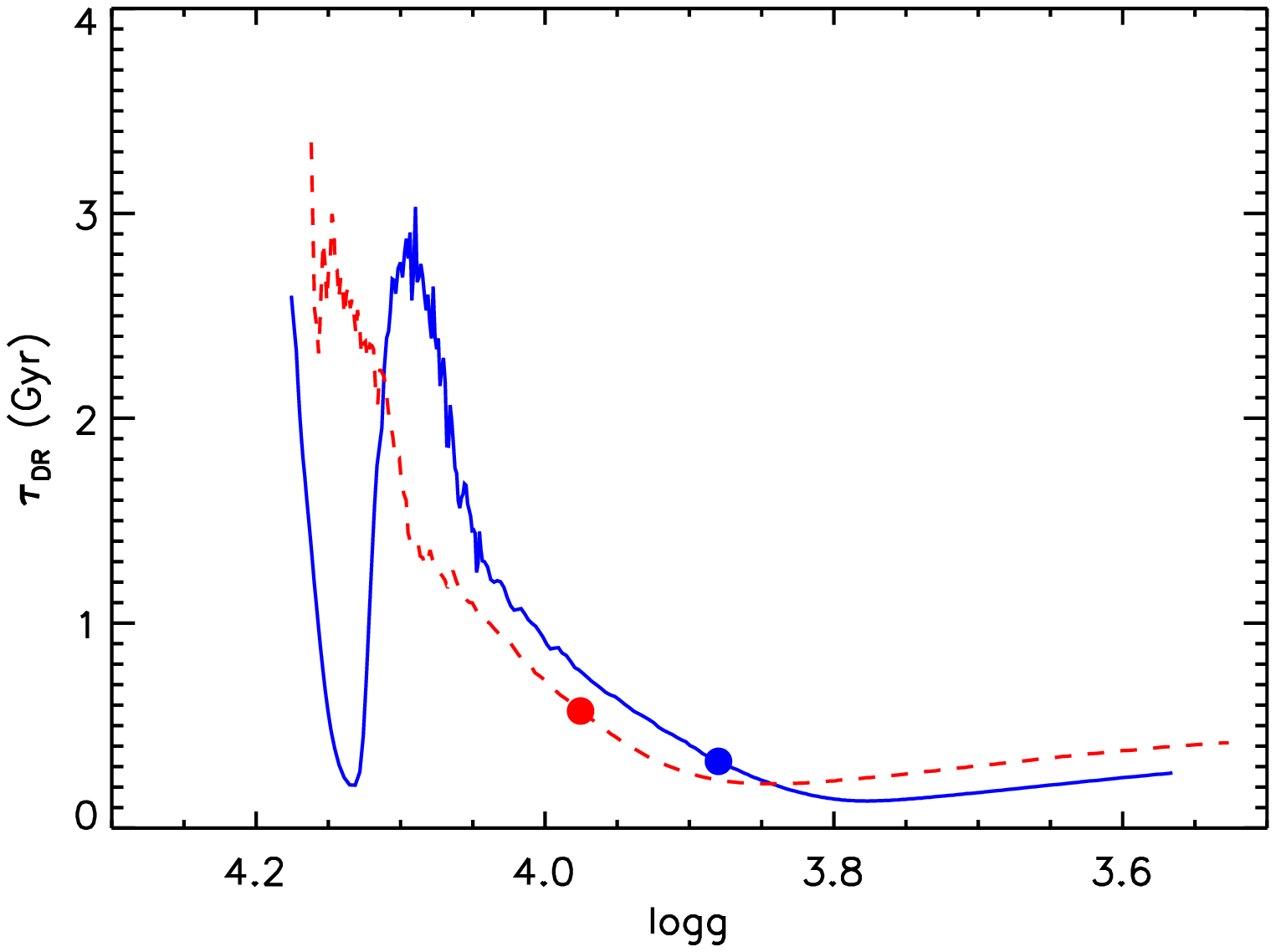}
\includegraphics[width=9cm]{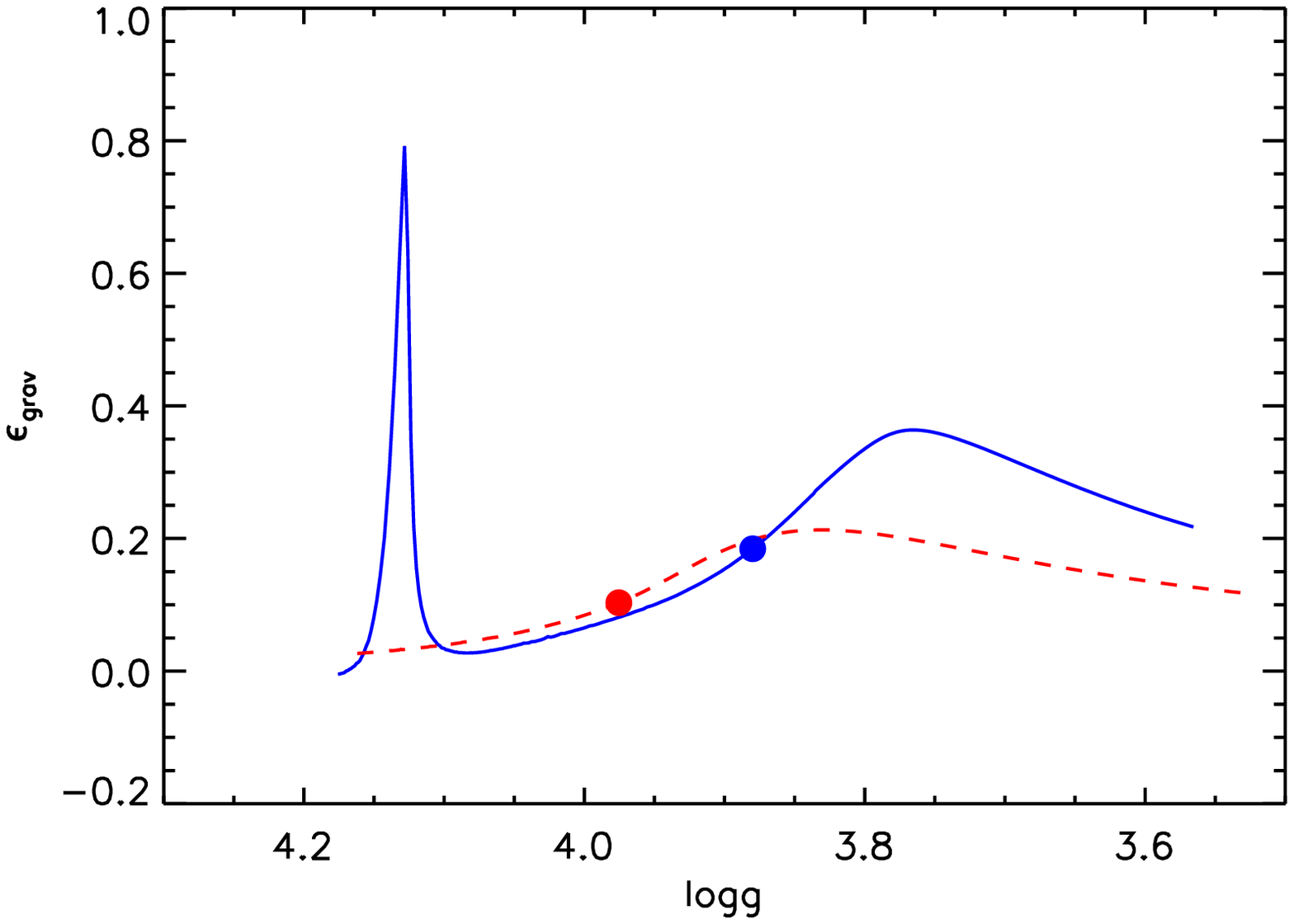}
\end{center}
\caption{Variations in the timescale $\tau_{\rm DR}$ over which differential rotation is forced (top panel) and in $\varepsilon_{\rm grav}$ (bottom panel) along the subgiant branch for our best-fit models of KIC8524425 (red dashed line) and KIC5955122 (solid blue line). The filled circles correspond to the stars at current age.
\label{fig_taudr}}
\end{figure}

\subsection{Forcing of differential rotation}

After the end of the main sequence, stars experience structural changes that are expected to force differential rotation. Without internal redistribution of AM, the contraction of the most central layers should spin up the core and the expansion of the envelope should make it spin down. To quantify the forcing of differential rotation for the two stars under study, we used the stellar models obtained in Sect. \ref{sect_seismic_model} to calculate the theoretical variations in the mean rotation rate $\langle\Omega\rangle_{\rm g}$ along the evolution, assuming that each layer conserves its specific AM. Under this hypothesis, we found that $\langle\Omega\rangle_{\rm g}$ should have increased by a factor of about four compared to its value at the end of the main sequence for both stars. By then, the envelope has expanded and thus spun down, so that we expect a core-envelope rotation contrast between eight and nine for the two stars. Therefore, an efficient transport of AM must take place to enforce nearly solid-body rotation, as was found from seismology.

\cite{eggenberger19} proposed to estimate the timescale of the forcing of differential rotation during the subgiant phase as $\tau_{\rm DR} = (\hbox{d}\ln\langle\Omega\rangle_{\rm g}/\hbox{d}t)^{-1}$, where $\langle\Omega\rangle_{\rm g}$ is computed assuming conservation of the specific AM. The top panel of Fig. \ref{fig_taudr} shows the variations in $\tau_{\rm DR}$ during the post main sequence evolution for both stars. Whatever mechanism redistributes AM in young subgiants has to operate on a timescale shorter than $\tau_{\rm DR}$ to enforce solid-body rotation. Two different phases where differential rotation is strongly forced were identified by \cite{eggenberger19}. The first one takes place just after the turnoff for stars that had a convective core during the main sequence (this is the case for KIC5955122, and Fig. \ref{fig_taudr} shows that $\tau_{\rm DR}$ drops to about 200 Myr after the MS turnoff). The second phase coincides with the base of the RGB, where $\tau_{\rm DR}$ decreases from about 3 Gyr to about 100 Myr for both stars. These rapid changes in $\tau_{\rm DR}$ can be explained by departures from thermal equilibrium, which are measured by $\varepsilon_{\rm grav}=-T\partial s/\partial t$. The bottom panel of Fig. \ref{fig_taudr} shows the variations in $\varepsilon_{\rm grav}$ in the stellar core of both stars, which are clearly anti-correlated with the variations in $\tau_{\rm DR}$. The forcing of differential rotation is maximal at the peaks of $\varepsilon_{\rm grav}$, that is when the core contracts on a thermal timescale. 

The stars studied by D14 all lie after the peak of $\varepsilon_{\rm grav}$ corresponding to the base of the RGB except for the least evolved one (star B), which lies precisely at the maximum of the peak (see Fig. 6 of \citealt{eggenberger19}). As shown by Fig. \ref{fig_taudr}, the evolutionary stage of KIC8524425 and KIC5955122 places them well before the maximum of  $\varepsilon_{\rm grav}$, which confirms that they are indeed less evolved than the sample of D14. The nearly solid-body rotation profiles that we obtained show that the internal redistribution of AM for these stars takes place on timescales lower than the current value of $\tau_{\rm DR}$ (about 320 Myr for KIC5955122 and 570 Myr for KIC8524425).

\subsection{Efficiency of AM transport}

To estimate the efficiency of the AM transport that is required to reproduce the observed rotation profiles, we computed rotating models for the two subgiants, including an additional constant viscosity $\nu_{\rm add}$ in the equation for the transport of AM in radiative zones. This approach was already followed for instance by \cite{eggenberger12}, \cite{spada16}, \cite{eggenberger19}, and \cite{denhartogh19}, for more evolved subgiants and red giants. As already mentioned in these works, this is a very crude description of the AM transport in subgiants. Naturally, we do not expect whatever mechanism is at work to indeed operate like a diffusive process with constant viscosity. The main goal of this approach is to study how the mean efficiency of the unknown transport processes vary with stellar parameters, and in particular with evolution. It was shown by \cite{eggenberger19} that the transport efficiency decreases during the late subgiant phase, near the base of the RGB, but must increase again as stars evolve along the RGB (\citealt{eggenberger17}). We here had the opportunity to estimate the efficiency of the transport closer to the MS turnoff and compare with its value on the MS. 

\subsubsection{Rotating models}

We used two different stellar evolution codes: \textsc{YREC} (\citealt{yrec}) and \textsc{GENEC} (\citealt{eggenberger08}). The two codes differ in their treatment of the rotation.

\textsc{GENEC} follows the assumption of shellular rotation advocated by \cite{zahn92} (see also \citealt{maeder98}). The evolution of the rotation profile is computed simultaneously to the evolution of the star, taking into account AM transport by meridional currents and the shear instability. In this context, the equation for AM transport that is solved in radiative regions is
\begin{equation}
\rho\frac{\hbox{d}}{\hbox{d}t}\left( r^2\Omega\right)_m = \frac{1}{5r^2} \frac{\partial}{\partial r} \left[ \rho r^4 \Omega U(r) \right] + \frac{1}{r^2} \frac{\partial}{\partial r} \left( \rho D_{\rm tot} r^4\frac{\partial\Omega}{\partial r} \right),
\label{eq_evolAM}
\end{equation}
where $\rho(r)$ is the density at radius $r$, and $U(r)$ corresponds to the radial dependence of the meridional circulation velocity in the radial direction. An effective diffusion coefficient $D_{\rm tot}$ was introduced as $D_{\rm tot}=D_{\rm shear}+\nu_{\rm add}$, where $D_{\rm shear}$ is the diffusion coefficient for AM transport by shear instability and $\nu_{\rm add}$ is the additional viscosity introduced. Models that include the braking of the stellar surface by magnetized winds were computed with the braking law of \cite{matt15} with a braking constant fixed to its solar-calibrated value. The initial rotation rates of the stars were adjusted in order to reproduce their observed current surface rotation rates.

For \textsc{YREC} models, we neglected the AM transport by meridional circulation and shear instabilities. We considered only the effects of the unknown mechanism that transports AM in subgiants, assuming that it acts as a diffusion process, as was done in \cite{spada16}. Thus, only the second term in the right-hand-side term of Eq. \ref{eq_evolAM} was retained, and $D_{\rm tot} = \nu_{\rm add}$. We did not expect this difference of treatment to affect much the estimates of $\nu_{\rm add}$ compared to the \textsc{GENEC} models because it has been shown that rotation-induced AM transport as it is currently understood is very inefficient in this phase of the evolution (e.g., \citealt{ceillier13}). The braking of the stellar surface through magnetized wind was included following the prescription of \cite{kawaler88}. The initial rotation period was chosen so that to the period at the age of 1 Myr is 8 days (this choice roughly coincides with the median of the observed period distribution of the Orion Nebula Cluster, \citealt{rebull01}). The braking constant $K$ of the wind model was then adjusted to match the surface rotation of the two stars.

\subsubsection{Results}

\begin{figure}
\begin{center}
\includegraphics[width=9cm]{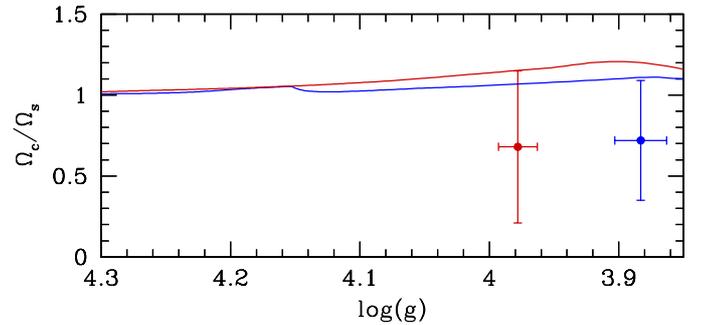}
\end{center}
\caption{Ratio between core and envelope rotation rates obtained with \textsc{GENEC} models computed with an additional viscosity $\nu_{\rm add}$ adjusted to match asteroseismic observations (see text). Filled circles indicate seismic measurements for KIC5955122 (blue symbol) and for KIC8524425 (red symbol).
\label{fig_nuadd}}
\end{figure}

Using both evolution codes, $\nu_{\rm add}$ was adjusted in order to reproduce the core to envelope rotation contrast obtained by seismology. Above a certain value of $\nu_{\rm add}$, the rotation profile is constant throughout the star. Since the rotation profiles of both subgiants were found to be consistent with solid-body rotation, we could only obtain a lower limit to the value of $\nu_{\rm add}$. Models computed with \textsc{GENEC} and \textsc{YREC} including magnetic braking during the MS yielded very similar results. They required $\nu_{\rm add} > 8\times10^4$ cm$^2$.s$^{-1}$ for KIC8524425 and $\nu_{\rm add} > 2.5\times 10^5$ cm$^2$.s$^{-1}$ for KIC5955122 to reproduce $\langle\Omega\rangle_{\rm g}/\langle\Omega\rangle_{\rm p}$ within 1-$\sigma$ errors (see Fig. \ref{fig_nuadd}). 

We also investigated how these results depend on the assumptions made concerning the transport of AM during the MS. For this purpose, we computed an additional set of \textsc{GENEC} models in the (unrealistic) case where no magnetic braking is included during the MS. For these models, slower initial rotations are needed to reproduce the current envelope rotation rates of the two stars. Naturally, these models reach the end of the MS with a lower level of radial differential rotation than the models computed with magnetized winds. We found that additional viscosity required is then lowered to about $5\times 10^4$ cm$^2$.s$^{-1}$ for KIC8524425, and $1.5\times 10^5$ cm$^2$.s$^{-1}$ for KIC5955122. We thus found that, contrary to more evolved stars, the rotational properties of young subgiants are sensitive to the internal transport of AM on the MS (\citealt{eggenberger19}).


\begin{figure}
\begin{center}
\includegraphics[width=9cm]{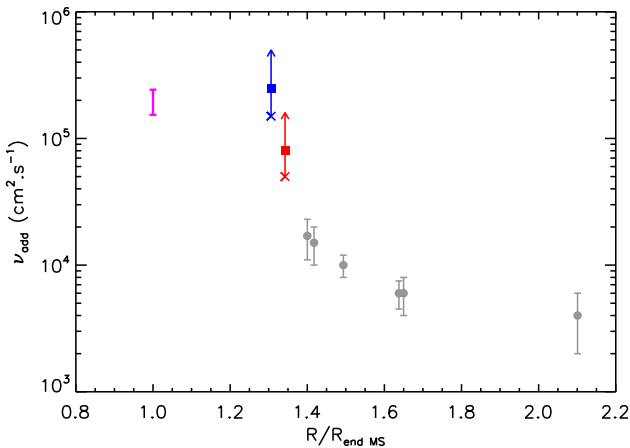}
\end{center}
\caption{Additional viscosity that is required to reproduce the seismic rotation profiles of subgiants as a function of the ratio between the current stellar radius and the radius at the end of the MS (proxy for evolutionary stage). Filled squares and vertical arrows indicate the lower limits on $\nu_{\rm add}$ obtained for KIC5955122 (blue) and KIC8524425 (red) with magnetic braking in the MS and crosses show the lower limits obtained without wind. Gray circles correspond to the evolved subgiants studied by \cite{eggenberger19}. The thick purple bar shows the efficiency of the AM transport during the main sequence for stars with masses between 1.1 and 1.2 $M_\odot$ obtained with open clusters (\citealt{spada19}).
\label{fig_evol_nuadd}}
\end{figure}

\subsubsection{Comparison with AM transport at other evolutionary stages}

The additional viscosities required for the two subgiants are significantly larger than those needed for the more evolved subgiants studied by D14 and \cite{eggenberger19} (whether or not we include magnetic braking during the main sequence). This confirms their findings, that the efficiency of the AM transport decreases during the subgiant phase. This is illustrated in Fig. \ref{fig_evol_nuadd}, where we have used the ratio between the current stellar radius and the radius at the MS turnoff as a proxy for the evolutionary stage (updated version of Fig. 7 of \citealt{eggenberger19}). 

The obtained values of $\nu_{\rm add}$ can also be compared to those obtained for MS stars using the surface rotation periods of stars in Galactic open clusters. These measurements can yield estimates of the timescale of the coupling between the core and the envelope in terms of AM transport as a function of the stellar mass. Recently, \cite{spada19} updated these estimates using observations of the Praesepe and NGC 6811 clusters. For stars with masses $\lesssim 1.3\,M_\odot$, they found  that the coupling timescale $\tau_{\rm c}$ varies as $\tau_{\rm c,\odot} (M/M_\odot)^{-\alpha}$, with $\tau_{\rm c,\odot} = 22\,\hbox{Myr}$ and $
\alpha = 5.6$. Using our mass estimates for KIC5955122 and KIC8524425 (see Table \ref{tab_model}), this yields MS coupling timescales of 7 and 12 Myr, respectively. \cite{denissenkov10a} showed that the two-zone model that was used by \cite{spada19} (among other authors) is equivalent to considering a diffusive transport of AM inside the star with a constant viscosity $\nu_{\rm add}$. They also provided an approximate correspondence between $\tau_{\rm c}$ and $\nu_{\rm add}$. Using this relation, we found that the MS coupling timescales for both stars translate into values of $\nu_{\rm add}$ in the range $(1.5$-$2.4)\times 10^5$ cm$^2$.s$^{-1}$. This interval is shown by the purple vertical bar in Fig. \ref{fig_evol_nuadd}. This shows that the efficiency of the additional mechanism that transports AM in young subgiants is found to be intermediate between the efficiency needed on the MS for these stars and the efficiency required later in the subgiant phase. This is a potential indication that the mechanism that transports AM during the MS might persist for some time after the end of the MS and become increasingly inefficient near the base of the RGB. 

\subsubsection{Comparison with predictions from potential AM transport mechanisms}


Internal gravity waves (IGW) excited at the bottom of the convective envelope were proposed as potential candidates for the transport of AM in subgiants. These waves propagate in the radiative interior, where they are damped and eventually dissipate. To produce a net transport of AM, a certain amount of radial differential rotation is needed. Indeed, with a flat rotation profile, the contributions from prograde and retrograde waves cancel each other out. \cite{pincon17} considered IGW excited by the penetration of turbulent plumes at the base of the convective envelope. They calculated the threshold of differential rotation $\Delta\Omega_{\rm th} = \Omega_{\rm core} - \Omega_{\rm env}$ above which the waves transport AM on a time scale shorter than the time scale of the core contraction. The authors suggested a self-regulating process, which imposes that the differential rotation stabilizes around the threshold value. \cite{pincon17} provided first estimates of $\Delta\Omega_{\rm th}$, which are in quite good agreement with the amount of differential rotation seismically inferred for young red giants by D14. Interestingly, \cite{pincon17} also found that for some time after the end of the MS, even a negligible amount of differential rotation is enough to couple the core to the envelope. For a 1.15-$M_\odot$ model (roughly matching the masses of our two subgiants), the radial differential rotation becomes detectable when $\log g \lesssim 3.93$ and this threshold in $\log g$ decreases with increasing stellar mass (see Fig. 6 of \citealt{pincon17}). Thus, although more detailed calculations are clearly required, it seems that AM transport by IGW could account for the nearly solid-body rotation profiles of KIC5955122 ($M\sim1.22\,M_\odot$ and $\log g\sim 3.88$) and KIC8524425 ($M\sim1.11\,M_\odot$ and $\log g\sim 3.98$). In any case, we recall that IGW cannot account on their own for the rotational evolution of more evolved red giants because the increasingly large \vaisala\ frequency in the core makes the transport of AM by waves more and more inefficient (\citealt{talon08}, \citealt{fuller14}).

\begin{figure}
\begin{center}
\includegraphics[width=9cm]{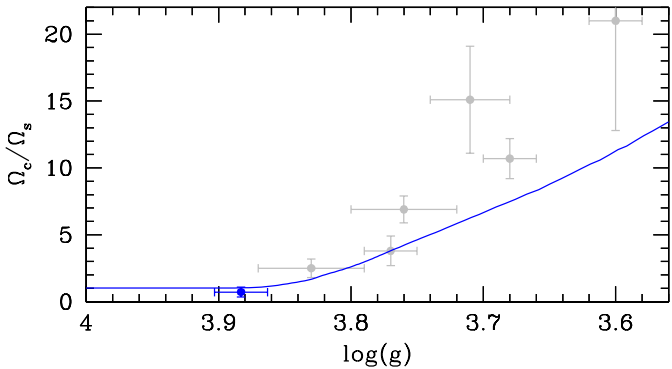}
\end{center}
\caption{Ratio between core and envelope rotation rates obtained with \textsc{GENEC} and the formalism of \cite{fuller19} with $\alpha=0.6$ (blue line). The filled blue circle corresponds to the observed core-envelope contrasts for KIC5955122. For visual comparison, the stars of D14 have been added to this plot (gray symbols). However, we note that evaluating the transport efficiency for these stars requires a dedicated modeling taking into account their masses and chemical compositions (see Fig. 3 of \citealt{eggenberger20}).
\label{fig_sg_fuller}}
\end{figure}

Internal magnetic fields could also be responsible for the transport of AM in subgiants. Recently, \cite{fuller19} proposed a revised prescription of the transport of AM by the Tayler instability. They considered an alternate saturation mechanism, which leads to larger magnetic field amplitudes at the saturation of the instability, and thus a more efficient transport of AM than the original prescription of \cite{spruit02}. They proposed revised expressions for the effective AM diffusivity linked to the Tayler instability and the minimum shear $q_{\rm min} = (-\partial\ln\Omega/\partial\ln r)_{\rm min}$ required to trigger the instability. They found that with this formalism, rigid rotation is expected to be maintained for some time after the end of the MS. The amount of differential rotation during the subgiant phase is essentially determined by the minimum shear $q_{\rm min}$, which depends on a dimensionless parameter $\alpha\sim 1$, according to the authors. \cite{eggenberger20} showed that values of $\alpha$ significantly below unity ($\alpha\sim0.5$) are needed to reproduce the internal rotation of the evolved subgiants studied by \cite{deheuvels14}, while matching the core rotation of stars on the RGB requires $\alpha\sim 1.5$. Moreover, even higher values of $\alpha$ are needed to account for the internal rotation of intermediate mass stars in the core He burning phase (\citealt{denhartogh20}). To estimate the agreement of this formalism with the results of this study, we computed a model of KIC5955122 including the prescription of \cite{fuller19} using the \textsc{GENEC} code. We found that values of $\alpha\gtrsim0.6$ are required to have the star rotate rigidly at its current age (see Fig. \ref{fig_sg_fuller}). The second subgiant, KIC8524425, is less constraining because it has a higher value of $\log g$, and so even relatively low values of $\alpha$ can produce near solid-body rotation for this star. This is consistent with the value inferred by \cite{eggenberger20} for the least evolved stars of D14, but yields a too low degree of differential rotation to account for the most evolved ones of the sample.

\section{Conclusion \label{sect_concl}}

In this study, we probed the internal rotation of two \kepler\ subgiants, KIC8524425 and KIC5955122, at an intermediate stage of evolution between the main sequence turnoff and the base of the red giant branch, where constraints on the core rotation were still lacking. We were able to measure eigenfrequencies and rotational splittings for dipole and quadrupole modes for these two stars. Through a detailed seismic modeling, we confirmed that the two subgiants are indeed closer to the main sequence turnoff than the stars studied by \cite{deheuvels14}. We also obtained rotational kernels for the detected modes and could therefore perform rotation inversions.

We obtained precise measurements of the average rotation period in the p-mode cavity (we found $\langle\Omega_{\rm p}\rangle/(2\pi)=298\pm20$~nHz for KIC8524425 and $675\pm27$~nHz for KIC5955122). These results are in very good agreement with the measurements of surface rotation rates obtained by \cite{garcia14} and \cite{bonanno14} for these two stars (within 1-$\sigma$ errors). The envelope rotation measured by seismology has a mean sensitivity around a normalized radius of 0.76 for both stars, that is very close to the bottom of the convective envelope for KIC5955122 ($r_{\rm CE}/R = 0.74$) and in the lower part of the convective envelope for KIC8524425 ($r_{\rm CE}/R = 0.65$). The close agreement between the seismic envelope rotation rate and the surface rotation rate thus shows that there can be only a mild radial differential rotation within the convective envelopes of these stars. This type of constraints could be helpful to investigate the nature of the mechanism that transports angular momentum (AM). Indeed, \cite{kissin15} have proposed that radial rotation gradients might be located within the convective envelope due to AM pumping, rather than in the radiative core. Measuring the rotation gradient in the convective envelope at various stages along the red giant evolution would enable us to test this statement.

Measuring the core rotation for the two subgiants was more challenging than for more evolved subgiants because we measured rotational splittings for few g-dominated modes. However, with the MOLA inversion technique, we were able to build averaging kernels that efficiently suppress the contribution from the envelope. We found $\langle\Omega_{\rm g}\rangle/(2\pi)=204\pm134$~nHz for KIC8524425 and $488\pm227$~nHz for KIC5955122. We thus derived core-envelope rotation ratios of $0.68\pm0.46$ and $0.72\pm0.37$ for KIC8524425 and KIC5955122, respectively. These results are consistent with a solid-body rotation for the two targets and they clearly show that the core cannot be rotating much faster than the envelope. Our results leave open the possibility that the core might be rotating slower than the envelope. Such a behavior has been reported before (\citealt{kurtz14}). It could not be produced by a mechanism of AM transport that would act to smooth out rotation gradients, such as a diffusion process. However, other mechanisms such as the transport of AM by internal gravity waves (IGW) can potentially lead to envelopes spinning faster than the core (\citealt{rogers13}).

The near solid-body rotation found for the two subgiants indicates that AM is transported faster than the timescale over which differential rotation is forced, which we found to be equal to about 320 Myr for KIC5955122, and 570 Myr for KIC8524425. To go further, we estimated the efficiency of the redistribution of AM by including an additional diffusion of AM with a constant, adjustable viscosity $\nu_{\rm add}$. We found that values of $\nu_{\rm add}>5\times10^4$~cm$^2$.s$^{-1}$ and $\nu_{\rm add}>1.5\times10^5$~cm$^2$.s$^{-1}$ are needed to reproduce the core-envelope contrast of KIC8524425 and KIC5955122, respectively, within 1-$\sigma$ errors. These values are higher than the additional viscosity that was required to account for the internal rotation of more evolved subgiants (\citealt{eggenberger19}), which brings further support to the claim of these authors, that the efficiency of the AM transport is decreasing during the subgiant phase. Also, the efficiency of the AM transport that is required for the two subgiants is lower than or comparable to the efficiency needed during the main sequence for stars of equivalent masses (\citealt{spada19}). This might indicate that the mechanism that flattens the rotation profile during the main sequence persists for some time after the exhaustion of hydrogen in the core.

Different mechanisms of AM transport are currently being investigated as candidates to account for the internal rotation of subgiants and red giants. We compared our results to the only two studies that have provided predictions for the evolution of rotation during the subgiant phase. \cite{pincon17} show that IGW excited by the penetration of plumes at the bottom of the convective envelope could produce efficient transport and they proposed that it might act as a self-regulating process. The near solid-body rotation of KIC8524425 and KIC5955122 seems to be roughly compatible with this scenario, although detailed calculations would be needed to check this. AM could also be transported through   instabilities of an internal magnetic field. This possibility remains to be fully explored. \cite{fuller19} proposed a revised prescription of the so-called Tayler-Spruit dynamo, which depends on a dimensionless parameter $\alpha$. Using their formalism, we found that the value of $\alpha$ that is required to match the core rotation of stars near the base of the RGB ($\alpha\sim0.6$) is compatible with a solid-body rotation of our two subgiants. However, as shown by \cite{eggenberger20} and \cite{denhartogh20}, a higher value of $\alpha$ ($\gtrsim 1.5$) is needed to reproduce the rotation along the RGB and in the secondary clump.

In this work, we have bridged the gap between the (few) constraints that we have on the internal rotation during the main sequence of solar-like pulsators and the constraints available from the base of the RGB. These new observations will need to be accounted for by the candidates for AM transport in red giants.

\begin{acknowledgements}
We thank the anonymous referee for comments that improved the clarity of the paper. S.D. and J.B. acknowledge support from the project BEAMING ANR-18-CE31- 0001 of the French National Research Agency (ANR) and from the Centre National d'Etudes Spatiales (CNES). P.E. has received funding from the European Research Council (ERC) under the European Union's Horizon 2020 research and innovation programme (grant agreement No 833925, project STAREX). JdH acknowledges support by the European Research Council (ERC-2016-CO Grant 724560).
\end{acknowledgements}

\bibliographystyle{aa.bst} 
\bibliography{biblio} 

\begin{appendix}

\section{Effects of latitudinal differential rotation on the splittings of quadrupole modes \label{app_a3}}

In Sect. \ref{sect_sismo}, we could not find significant evidence for latitudinal differential rotation in the convective envelope of the two stars under study. However, we did obtain constraints on the $a_3$ coefficients. For KIC5955122, we checked whether this measurement is consistent with the photometric estimates of the latitudinal differential rotation at the stellar surface obtained by \cite{bonanno14}. These authors found an equatorial rotation period of 16.4 days and estimated the difference between the rotation rate at the pole $\Omega_{\rm pole}$ and the rotation rate at the equator $\Omega_{\rm eq}$  to $\Delta\Omega \equiv \Omega_{\rm pole} - \Omega_{\rm eq} = -0.25\pm0.02$ rad.d$^{-1}$.

To estimate the expected effect of latitudinal differential rotation on the mode splittings, we parametrized the rotation profile $\Omega(r,\theta)$ as
\begin{equation}
    \Omega(r,\theta) = \Omega_0(r)\psi_0(\theta) + \Omega_1(r)\psi_1(\theta).
    \label{eq_om2D}
\end{equation}
The function $\Omega_0(r)$ gives a measurement of the radial differential rotation within the star, while $\Omega_1(r)$ measures the latitudinal differential rotation. We here assumed that the latitudinal differential rotation is restricted to the convective envelope and independent of the radius in this region. We thus chose
\begin{equation}
    \Omega_1(r) = \left\{
                \begin{array}{ll}
                  \Omega_1 & \hbox{if} \; r\geqslant r_{\rm CE}\\
                  0 & \hbox{otherwise,} 
                \end{array}
              \right.
\end{equation}
where $r_{\rm CE}$ is the radius of the bottom of the convective envelope. In this study, we found that KIC5955122 has negligible radial differential rotation (see Sect. \ref{sect_inversions}). Accordingly, we here considered $\Omega_0$ to be constant throughout the star.
The advantage of using orthogonal polynomials in the decomposition of Eq. \ref{eq_ritzwoller} is that the functions $\Omega_{s}(r)$ can be determined from the coefficients $a_{2s+1}^{(l)(n)}$ alone when the $\psi_{s}(\theta)$ correspond to a well chosen set of polynomials. For $s=0,1$, (see, e.g., \citealt{gizon04}) they correspond to
\begin{equation}
    \psi_0(\theta) = 1 \;\; ; \;\; \psi_1(\theta) = \frac{3}{2}(5\cos^2\theta-1).
    \label{eq_psi}
\end{equation}

The values of $\Omega_0$ and $\Omega_1$ are determined in a unique way by the measurements of the surface rotation given by \cite{bonanno14}, which yield
\begin{align}
    \Omega_0 & = \Omega_{\rm eq} + \frac{\Delta\Omega}{5} \\
    \Omega_1 & = \frac{2}{15}\Delta\Omega.
\end{align}

The $a_3^{(2)}$ coefficient is then given by
\begin{equation}
    a_3^{(2)}  = \frac{\Omega_1}{5}\int_{r_{\rm CE}}^R\int_0^\pi [K_{n,2,2}(r,\theta) - K_{n,2,1}(r,\theta)] \psi_1(\theta) \,\hbox{d}\theta\hbox{d}r.
    \label{eq_a3}
\end{equation}
The rotational kernels of the modes $K_{n,l,m}(r,\theta)$, whose expressions are given for instance by \citealt{gizon04}), were obtained using our best-fit model from Sect. \ref{sect_seismic_model}. Using Eq. \ref{eq_a3}, we found that the measurements by \cite{bonanno14} correspond to an $a_3^{(2)}$ coefficient of about 16~nHz. This value is consistent with our seismic measurement of $a_3^{(2)} = 16^{+23}_{-25}$~nHz for KIC5955122.

\section{Estimated mode frequencies and rotational splittings \label{app_seismic_params}}

Tables \ref{tab_852} and \ref{tab_595} give the oscillation mode parameters that were obtained for KIC8524425 and KIC5955122, respectively, in Sect. \ref{sect_sismo}.



\begin{table}
  \begin{center}
  \caption{Estimated mode frequencies and rotational splittings (only for modes that were found to be significantly split by rotation, see Sect. \ref{sect_sismo}) for KIC8524425.
  \label{tab_852}}
\vspace{0.2cm}
\begin{tabular}{c c c}
\hline \hline
\T \B $l$ & $\nu$ ($\mu$Hz) & $\delta\nu_{\rm s}$ ($\mu$Hz) \\
\hline
\T \B  0 & $ 743.703^{+0.062}_{-0.052}$ & n.a. \\
\T \B  0 & $ 803.166^{+0.106}_{-0.106}$ & n.a. \\
\T \B  0 & $ 861.175^{+0.052}_{-0.053}$ & n.a. \\
\T \B  0 & $ 919.584^{+0.034}_{-0.033}$ & n.a. \\
\T \B  0 & $ 978.931^{+0.034}_{-0.034}$ & n.a. \\
\T \B  0 & $1038.655^{+0.027}_{-0.027}$ & n.a. \\
\T \B  0 & $1098.319^{+0.029}_{-0.029}$ & n.a. \\
\T \B  0 & $1158.056^{+0.050}_{-0.050}$ & n.a. \\
\T \B  0 & $1217.934^{+0.079}_{-0.080}$ & n.a. \\
\T \B  0 & $1278.321^{+0.167}_{-0.167}$ & n.a. \\
\hline
\T \B  1 & $ 782.339^{+0.055}_{-0.061}$ & $   0.205^{+0.064}_{-0.107}$ \\
\T \B  1 & $ 832.875^{+0.052}_{-0.049}$ & $   0.352^{+0.050}_{-0.054}$ \\
\T \B  1 & $ 888.544^{+0.039}_{-0.040}$ & $   0.322^{+0.047}_{-0.051}$ \\
\T \B  1 & $ 945.322^{+0.040}_{-0.040}$ & - \\
\T \B  1 & $1001.819^{+0.037}_{-0.036}$ & $   0.301^{+0.057}_{-0.072}$ \\
\T \B  1 & $1044.829^{+0.021}_{-0.019}$ & $   0.207^{+0.024}_{-0.025}$ \\
\T \B  1 & $1073.973^{+0.026}_{-0.026}$ & $   0.292^{+0.049}_{-0.041}$ \\
\T \B  1 & $1127.333^{+0.032}_{-0.031}$ & $   0.304^{+0.067}_{-0.066}$ \\
\T \B  1 & $1185.438^{+0.053}_{-0.053}$ & - \\
\T \B  1 & $1245.027^{+0.072}_{-0.074}$ & - \\
\T \B  1 & $1305.243^{+0.155}_{-0.155}$ & - \\
\T \B  1 & $1365.151^{+0.219}_{-0.219}$ & - \\
\hline
\T \B  2 & $ 797.272^{+0.369}_{-0.324}$ & - \\
\T \B  2 & $ 856.537^{+0.109}_{-0.111}$ & $   0.192^{+0.080}_{-0.091}$ \\
\T \B  2 & $ 914.464^{+0.071}_{-0.064}$ & $   0.313^{+0.056}_{-0.046}$ \\
\T \B  2 & $ 973.679^{+0.069}_{-0.069}$ & $   0.266^{+0.048}_{-0.047}$ \\
\T \B  2 & $1032.794^{+0.055}_{-0.048}$ & $   0.334^{+0.049}_{-0.042}$ \\
\T \B  2 & $1093.455^{+0.042}_{-0.043}$ & $   0.238^{+0.038}_{-0.035}$ \\
\T \B  2 & $1153.064^{+0.066}_{-0.064}$ & - \\
\T \B  2 & $1212.760^{+0.119}_{-0.111}$ & - \\
\T \B  2 & $1273.530^{+0.190}_{-0.197}$ & - \\
\hline
\hline
\multicolumn{3}{l}{$^a$: not applicable}
\end{tabular}
\vspace{0.5cm}
\end{center}
\end{table}

\begin{table}
  \begin{center}
  \caption{Same as Table \ref{tab_852} for KIC5955122.
  \label{tab_595}}
\vspace{0.2cm}
\begin{tabular}{c c c}
\hline \hline
\T \B l & $\nu$ ($\mu$Hz) & $\delta\nu_{\rm s}$ ($\mu$Hz) \\
\hline
\T \B  0 & $ 560.765^{+0.108}_{-0.115}$ & n.a. \\
\T \B  0 & $ 609.570^{+0.140}_{-0.215}$ & n.a. \\
\T \B  0 & $ 658.087^{+0.318}_{-0.321}$ & n.a. \\
\T \B  0 & $ 706.319^{+0.100}_{-0.101}$ & n.a. \\
\T \B  0 & $ 754.504^{+0.056}_{-0.058}$ & n.a. \\
\T \B  0 & $ 803.811^{+0.044}_{-0.046}$ & n.a. \\
\T \B  0 & $ 853.529^{+0.043}_{-0.044}$ & n.a. \\
\T \B  0 & $ 903.083^{+0.065}_{-0.065}$ & n.a. \\
\T \B  0 & $ 952.477^{+0.096}_{-0.095}$ & n.a. \\
\T \B  0 & $1002.729^{+0.150}_{-0.165}$ & n.a. \\
\T \B  0 & $1053.372^{+0.487}_{-0.515}$ & n.a. \\
\T \B  0 & $1103.505^{+0.205}_{-0.107}$ & n.a. \\
\hline
\T \B  1 & $ 586.316^{+0.143}_{-0.105}$ & $   0.498^{+0.087}_{-0.139}$ \\
\T \B  1 & $ 623.600^{+0.077}_{-0.081}$ & $   0.643^{+0.067}_{-0.071}$ \\
\T \B  1 & $ 657.738^{+0.338}_{-0.905}$ & - \\
\T \B  1 & $ 690.107^{+0.057}_{-0.057}$ & $   0.627^{+0.067}_{-0.076}$ \\
\T \B  1 & $ 731.201^{+0.053}_{-0.054}$ & $   0.534^{+0.088}_{-0.118}$ \\
\T \B  1 & $ 774.722^{+0.045}_{-0.046}$ & $   0.735^{+0.052}_{-0.053}$ \\
\T \B  1 & $ 809.721^{+0.037}_{-0.036}$ & $   0.442^{+0.045}_{-0.049}$ \\
\T \B  1 & $ 836.690^{+0.040}_{-0.040}$ & $   0.574^{+0.051}_{-0.052}$ \\
\T \B  1 & $ 879.858^{+0.053}_{-0.055}$ & $   0.636^{+0.097}_{-0.119}$ \\
\T \B  1 & $ 927.408^{+0.054}_{-0.058}$ & - \\
\T \B  1 & $ 975.989^{+0.078}_{-0.079}$ & - \\
\T \B  1 & $1024.921^{+0.120}_{-0.121}$ & - \\
\T \B  1 & $1073.550^{+0.182}_{-0.188}$ & - \\
\hline
\T \B  2 & $ 653.409^{+0.317}_{-0.303}$ & $   0.563^{+0.166}_{-0.213}$ \\
\T \B  2 & $ 702.052^{+0.270}_{-0.276}$ & $   0.675^{+0.138}_{-0.162}$ \\
\T \B  2 & $ 749.926^{+0.106}_{-0.109}$ & $   0.634^{+0.060}_{-0.060}$ \\
\T \B  2 & $ 799.594^{+0.101}_{-0.097}$ & $   0.703^{+0.071}_{-0.058}$ \\
\T \B  2 & $ 849.202^{+0.086}_{-0.087}$ & $   0.623^{+0.049}_{-0.050}$ \\
\T \B  2 & $ 898.478^{+0.122}_{-0.124}$ & $   0.669^{+0.072}_{-0.073}$ \\
\T \B  2 & $ 948.349^{+0.130}_{-0.128}$ & - \\
\T \B  2 & $ 998.270^{+0.209}_{-0.217}$ & - \\
\T \B  2 & $1048.951^{+0.511}_{-0.518}$ & - \\
\hline
\hline
\end{tabular}
\vspace{0.5cm}
\end{center}
\end{table}

\end{appendix}

\end{document}